\documentclass[11pt]{article}
\usepackage{graphicx}
\usepackage{natbib}
\bibpunct{(}{)}{;}{a}{}{,} 
\let\jnlstyle=\rm
\def\refjnl#1{{\jnlstyle#1\,}}

\def\apj{\refjnl{Astrophys.J.}}

\def\mnras{\refjnl{Mon.Not.Roy.astron.Soc.}}
\def\memras{\refjnl{Mem.Roy.astron.Soc.}}
\def\prc{\refjnl{Phys.~Rev.~C}}

\def\pre{\refjnl{Phys.~Rev.~E}}
\def\physrep{\refjnl{Phys.~Rep.}}

\def\apjs{\refjnl{Astrophys.J. Suppl.}}

\def\aap{\refjnl{Astron.Astrophys.}}

\begin{document}


\title{The equation of state and composition of hot, dense matter in core-collapse supernovae.
     }
  \author{S.I. Blinnikov$^{1,2,3}$,  I.V. Panov$^2$, M.A.Rudzsky$^4$, K. Sumiyoshi$^5$ \\
\\
$^1${Max-Planck-Institut f\"ur Astrophysik,}\\
{Karl-Schwarzschild-Strasse 1, Postfach 1523, D-85740 Garching, Germany}\\
\and
 $^2$     Institute for Theoretical and experimental Physics,  \\
B.~Cheremushkinskaya St.~25, 117218 Moscow, Russia\\
\and
$^3$  IPMU, University of Tokyo
5-1-5 Kashiwanoha, Kashiwa, 277-8568, Japan \\
\and
\\
$^4$ Computer Science Department, Technion-Israel Institute of Technology,\\
32000, Haifa, Israel \\
\and
$^5${ Numazu College of Technology, Ooka 3600, Numazu, Shizuoka 410-8501, Japan}
}

\date{}

\maketitle

\flushright{Preprint IPMU~09-0045}



\par\medskip\noindent
%

%


\abstract{
The equation of state and composition of matter are calculated
for conditions typical for pre-collapse and early collapse stages
in core collapse supernovae. The composition is evaluated under the assumption
of nuclear statistical equilibrium, when the matter is
considered as an `almost' ideal gas with corrections due to thermal excitations
of nuclei, to free nucleon degeneracy, and to Coulomb and surface energy
corrections. The account of these corrections allows us to obtain
the composition for densities a bit below the nuclear matter density.
Through comparisons with the equation of state (EOS) developed by Shen et al.
we examine the approximation of one representative nucleus used in
most of recent supernova EOS's.
We find that widely distributed compositions in the nuclear chart
are different, depending on the mass formula,
while the thermodynamical quantities are quite close to those in the Shen's EOS.
}

{\bf keywords:} {Nuclear reactions, nucleosynthesis, abundances --
                Stars: neutron -- supernovae: general
               }


\section{Introduction}

     A proper description of the equation of state (EOS)
for subnuclear and supernuclear
densities is of vital importance for current studies on the explosion
mechanism of core-collapse supernovae \citep{bet90,suz94,jan07}.
The success of the prompt shock propagation in the supernova mechanism
depends on the size of the homologously  contracting inner part of a
collapsing   core.  The larger is $Y_{\rm e}$ (the number of electrons per
nucleon)
at the time of bounce, the smaller part of  matter will be dissociated
and the stronger will be the prompt shock wave.
Therefore, all factors which influence  $Y_{\rm e}$  must be thoroughly studied.

One factor is the mass fraction of the free protons, $X_{\rm p}$,
in the dense matter of supernova core.
Since the rate of electron capture on a free proton is much higher
than that on a nucleus, free protons play an important role in establishing
the value of $Y_{\rm e}$ at the neutrino trapping.
The abundance of free protons depends sensitively on the nuclear models
of dense matter.
The variations of $X_{\rm p}$ amount to more than order of magnitude
in the studies of supernova EOS
so far \citep{coo85,coo90,hil85,hil91,lat85,lat91,she98a,she98b} and
can affect the initial strength of shock wave \citep{bru89a,swe94,sum04,sum05}.
As shown by \citet{bru89a,bru89b}, the free proton fraction may
be largely different depending on the nuclear interaction and its model.
Another factor is the composition of various nuclei, which exist
under the nuclear statistical equilibrium.
The total rate of electron captures on nuclei can dominate that
on free protons, if the number of the free protons is small \citep{hix03},
and nuclei may affect the dynamics in this way as well.
Therefore, it is crucial to evaluate the composition of dense matter
in a precise manner.

Another astrophysical problem, where an accurate chemical composition
is important, is the nucleosynthesis of heavy elements.
The studies of various models of the explosive nucleosynthesis
as well as the rapid neutron capture process
 \citep[see, e.g.][and references therein]{pti82,woo92,nad98,arn07}
require knowledge of the chemical composition 
close to
the nuclear statistical equilibrium (NSE) as an initial condition
for nuclear reaction network calculations.
Neutron star mergers and collapsar models for gamma-ray bursts
may contain similar conditions as well.

The thermodynamic properties of hot, dense matter have been investigated
in various approaches: Saha-like equations \citep{maz79,ele80,mur80,ish03,nad04},
Hartree-Fock approaches \citep{bon81,wol83,hil85,las87},
compressible liquid-drop model \citep{bay71,lat85,lat91},
and the relativistic mean field theory \citep{sut99}
with local density approximation \citep{she98a,she98b}.
The review of those approaches, beginning from \citet{BBS1970},
can be found in the context of cold neutron star matter in  \citet{Ruester06} and \citet{HPYa07}.
In most of the sets of EOS used for recent supernova simulations,
the approximation of one species of nuclei (in addition to
neutrons, protons and alpha particles) has been adopted \citep{hil85,lat91,she98b}.
However, the advance in recent studies of electron capture rates
on nuclei for a wide mass range \citep{lan03}
necessitates the evaluation of multi-composition of nuclei
to determine the total electron capture rate.
Although the treatment with one nuclear species should be a good
approximation to derive overall thermodynamical properties
of dense matter \citep{bur84}, 
it is necessary to take into account the multi-composition
at high densities $\sim$10$^{12}$ g/cm$^3$
near the neutrino trapping regime, where nuclear interaction
and multi-compositional treatment may become
more influential to determine the equation of state.
The description of the EOS with the multi-composition is,
of course, crucial to predict the abundance
of nuclei and free protons, see e.g. \citet{mur80}.

In spite of an extensive use of the Saha equation in the literature,
not all relevant physics was included in the published work within
this approach from the very beginning.
For example,
\citet{mur80} has not taken into account the effects of non-ideal nucleon gas.
\citet{ele80} have consistently taken
into account finite temperature effects in nucleon interactions
following \citet{ele77},
but they have omitted the Coulomb corrections.
Later \citet{hil81}, \citet{BraGar99} and \citet{nad05} included them
in different approximations and solved the implicit set of Saha equations.
Moreover,  \citet{hil81} have taken into account an `excluded-volume' effect due to the
finite size of nuclei.
It was noted by \citet{hil91} that
the number of nuclear species included in the Saha treatment
strongly affects the mass fraction of free protons, which may change
the effective adiabatic index through electron captures.

In this paper we describe our code suitable for studying the properties of dense matter
relevant to core-collapsing supernova conditions, and we put an emphasis on
the composition of nuclei.
Our code is originally rather old: we have compared our NSE results with the kinetic approach already in the paper  \citep{pan01}, but the code has not been described in detail.
Another motivation for our work is the availability of new mass formulae
like \citep{kou05} and \citep{mol95} and detailed EOS tables like \citep{she98b}.
Our aim is to develop and to test a practical and reliable tool for predictions of NSE
composition at subnuclear densities.

Following most closely the conventional Saha approach \citep{CliTay65} and the method
by \citet{maz79},
we extend this approach in the following points:

\begin{itemize}
\item{} We take into account, at a certain level of approximation,
the influence of free nucleon gas on the surface and Coulomb energies
of nuclei.  We retain some terms which were omitted by \citet{maz79}.
However, we do not take into account the `excluded-volume' effect, not pretending
to reach very high densities.
\item{} We have an option to include various results for nuclear partition functions
like those of \citet{fow78}
and newer partition functions by \citet{eng91}
and \citet{eng90}.
\item{} Our network is appreciably larger than the one used previously.
The atomic mass table is updated using recent theoretical compilations
of atomic masses.
It covers $\sim$20000 nuclides \citep{kou07} for the KTUY mass formula \citep{kou05}
and $\sim$9000 nuclides for the FRDM mass formula \citep{mol95}
as an extra option.

\end{itemize}

To examine the basic properties of the equation of state for supernovae
in the current formulation,
we calculate the properties of dense matter covering a wide range of density,
$\rho$, electron fraction, $Y_{\rm e}$, and temperature, $T$.
We report here the results at the equilibrium for the nuclear and electromagnetic
processes, i.e. the NSE, for fixed values of $Y_{\rm e}$ without imposing beta
equilibrium.
The beta-equilibrium is easily incorporated in the Saha approach
as in \citet{maz79} and \citet{ele80}, and our code has such an option.
A brief discussion of account of the beta equilibrium is also given below.
We show the global features of the dense matter (thermodynamical
quantities and compositions) for the supernova environment.
In order to assess the dependence of 
the nuclear mass, we make the comparisons among several choices.
We also examine the similarity and difference from the single
nuclear species treatment by comparisons
with the Shen EOS table \citep{she98b}.

We explain the formulation of nuclear statistical equilibrium
in \S \ref{sec:incl} with the detailed description of spin of nuclei
in \S \ref{sec:spin}.
We describe briefly the atomic mass data used for the calculations
in \S \ref{sec:mass}, the Shen equation of state in \S \ref{sec:shen}
and the beta equilibrium in \S \ref{sec:beta}.
We show the properties of dense matter in the current framework
in \S \ref{sec:result}.
The summary and discussions will be given in \S \ref{sec:summary}.

\section{Thermodynamics of interacting nuclides}
\subsection{Inclusion of nuclear and Coulomb contributions}
\label{sec:incl}

The basis is the equation of nuclear statistical equilibrium
(NSE) with respect to strong and electromagnetic processes
for the chemical potentials of nuclei:
\begin{eqnarray}
\mu_i=Z_i\mu_{\rm p}+(A_i-Z_i)\mu_{\rm n} \;.
\label{eq:ravn}
\end{eqnarray}
We denote $i=1=p$ for proton and $i=2=n$ for neutron.
We use the index $i$ mostly for nuclei $(A_i,Z_i)$ with $A>1$,
but we include nucleons ($i=1, 2$) in the summation over all species.
To find the correct expressions
for $\mu_i$ we suppose that the free energy of the set of
$\{N_i\}$ nuclides is
\begin{eqnarray}
F_{\rm nuc}=\sum_i N_i\Phi_i(T,\{n_k\})\;.
\label{eq:helmen}
\end{eqnarray}
Here $\Phi_i$ is the free energy per nucleus $i$ and it may depend
not only on the concentration,
\begin{eqnarray}
n_i=N_i/V \;,
\label{eq:concen}
\end{eqnarray}
of the $i$-th nucleus but also on concentrations of some other nuclei
$n_k$. Then we have by definition
\begin{eqnarray}
\mu_i={\partial F_{\rm nuc}\over \partial N_i}\bigg |_{T,V}
=\Phi_i +\sum_k n_k{\partial \Phi_k\over \partial n_i}\bigg |_T\;.
\label{eq:chempo}
\end{eqnarray}
Thus we find for the pressure
\begin{eqnarray}
P_{\rm nuc}=-{\partial F_{\rm nuc}\over \partial V}\bigg |_{T,\{N_k\}}=
\sum_k\mu_kn_k -F_{\rm nuc}/V
\label{eq:press}
\end{eqnarray}
which is consistent with the definition of the grand thermodynamic
potential, $PV$. We have to give these elementary details here
since we find that some terms are omitted in the expressions for
$\mu_i$, e.g. by \citet{yak89}.

For example, let us start with the Coulomb contribution to
$F_{\rm nuc}$. As usual we introduce
\begin{eqnarray}
\Gamma_i ={1\over k_{\rm B}T}\biggl(
{e^2Z_i^2\over a_{Z_i}}\biggr) ={e^2Z_i^2\over k_{\rm B}T} \biggl(
{4\pi \over 3} n_i\biggr)^{1/3}
\label{eq:pusk}
\end{eqnarray}
with $a_{Z_i}$ being the radius of the ion sphere:
\begin{eqnarray}
{4\pi \over 3}a_{Z_i}^3=n_i^{-1}\;.
\label{eq:ionsph}
\end{eqnarray}
If $Q_i$ denotes the Coulomb correction to the free energy
and has the functional form
\begin{eqnarray}
 Q_i \equiv k_BTf_0(\Gamma_i)
\label{eq:zanoza}
\end{eqnarray}
then we find the Coulomb contribution to the chemical potential as
\begin{eqnarray}
\mu_i^{\rm Coul}=k_BT\biggl[f_0(\Gamma_i)+{1\over 3}{\partial f_0 \over
\partial \ln \Gamma_i}\biggr]
\label{eq:gusj}
\end{eqnarray}
contrary to Eq.(17) of \citet{maz79} 
and to Eq.(97) of \citet{yak89} 
who find $\mu_i=k_BT f_0(\Gamma_i)$,
but their expressions lead to zero correction to $P_{\rm nuc}$ in Eq. (\ref{eq:press}).
On the other hand, our expression Eq. (\ref{eq:gusj}) gives the expected correction to
$P_{\rm nuc}$ after substitution into Eq. (\ref{eq:press}).
Basically the formulae in \citet{yak89}  are correct, since one is
free to separate Coulomb contributions, ascribing them
either to electrons or ions, but one has to avoid confusion
with this separation.
In the book by \citet{HPYa07} they mostly rely on free energy $F$ instead of $\mu_i$.
Actually, the expression which is equivalent to (\ref{eq:gusj}) appeared already in \cite{Dewitt73},
however,
the argument on the role  of the chemical potentials and debates on approaches to them
still persists. For example,  \citet{Brown06} discuss the difference of their use of chemical potentials
with that of \citet{Dewitt73}. Essentially,  \citet{Brown06} found that the
Coulomb correction found by \citet{Dewitt73} for a classical plasma is applicable to
quantum plasma as well.
 \citep[See also][for a detailed review on modern perspective on Coulomb
systems]{Brydges99}.

We closely follow \citet{maz79}
in the assumed form of the free energy but with some modifications.
The main results presented below use the Coulomb correction as in Eq.(\ref{eq:zanoza}), as was
also taken by  \citet{maz79}. In the end we check the effect of the additional  term in
Eq.(\ref{eq:gusj}).
Thus, we put
\begin{eqnarray}
   \Phi_{\rm p}=\Phi_{\rm p}^{(0)}+W+Q_{\rm p}+B_{\rm p} \; ,
\label{eq:prosha}
\end{eqnarray}
\begin{eqnarray}
   \Phi_{\rm n}=\Phi_{\rm n}^{(0)}+W+B_{\rm n}     \; .
\label{eq:nosha}
\end{eqnarray}
Here $\Phi_{\rm p}^{(0)}$ and $\Phi_{\rm n}^{(0)}$
are the Fermi-Dirac expressions for
the free energy of non-relativistic nucleons and they are expressed
through the Fermi integrals of half-integer index.
To calculate them we use the code by \citet{nad74a,nad74b}.
(The superscript zero indicates that an expression is for
non-interacting particles). For the bulk interaction energy of free
nucleons we take the zero temperature expression from \citet{maz79}
where they have used the expression of \citet{bay71}
with minor corrections from \citet{mac76}:
\begin{eqnarray}
   W=W(k,\beta)
\label{eq:aleks}
\end{eqnarray}
with $k^3=1.5\,\pi^2(n_{\rm p}+n_{\rm n})$ and
$\beta=n_{\rm p}/(n_{\rm p}+n_{\rm n})$. We have used the same values
of parameters entering in Eq. (\ref{eq:aleks}) as \citet{maz79}:
$k_0=1.34 {\rm ~fm}^{-1}$; $W_0=15.5$ MeV; $s=27.1$ MeV; $K=268$ MeV.

For nuclei ($i>2$) we assume:
\begin{eqnarray}
   \Phi_i=\Phi_i^{(0)}+Q_i+B_i \; .
\label{eq:jasha}
\end{eqnarray}
The binding energy in Eqs. (\ref{eq:prosha}), (\ref{eq:nosha})
and (\ref{eq:jasha}) is
\begin{eqnarray}
    B_i=M_i-{ A_i \over A_m }M_m  \; ,
\label{eq:bengy}
\end{eqnarray}
with $M_i$ being the mass of a nucleus number $i$, which depends on the
choice of the reference nucleus $m$. We use $^{56}$Fe as the reference
nucleus in our calculations. In calculating the total energy
we exclude electron mass $m_{\rm e}$ from the electron energy, then
we have
\begin{eqnarray}
    B_i=\Delta M_i - { A_i \over A_m }\Delta M_m \; .
\label{eq:bendel}
\end{eqnarray}
where $\Delta M_i$ is the atomic mass excess, which accounts for
the mass of electrons.
To compare with \citet{lat91}
one has to take into account
that they assume
$B_{\rm n}=0$, where $m=2={\rm n}$ in our notation, and
$B_i=\Delta M_i - A_i\Delta M_{\rm n}$.

  Considering the Coulomb corrections we follow \citet{lat85} 
and take for the total Coulomb energy of a nucleus
\begin{eqnarray}
      E_i^{Coul}={3 \over 5} { Z_i^2 e^2 \over r_A }
           \bigl[1- {3 \over 2} {r_A \over a_Z} + {1\over 2}
               \bigl({r_A \over a_Z}\bigr)^3\bigr] \; .
\label{eq:coulen}
\end{eqnarray}
Here $r_A$ is the nuclear radius, $r_A=1.2\cdot10^{-13}A^{1/3}$ cm.
So the Coulomb correction is
\begin{eqnarray}
Q_i={3 \over 5} { Z_i^2 e^2 \over r_A }
\bigl[-{3 \over 2} {r_A \over a_Z} + {1\over 2}
               \bigl({r_A \over a_Z}\bigr)^3\bigr] \; .
\label{eq:nachalo}
\end{eqnarray}
In the leading term, which is what we need at high density, this is just $-0.9\,\Gamma_i$
with $\Gamma_i$ defined in
Eq.(\ref{eq:pusk}) when the form (\ref{eq:zanoza}) is used.
With the modification as in Eq.(\ref{eq:gusj}) it will be $-1.2\,\Gamma_i$ -- very close to the numbers
in Eqs. (56) and (57) found by  \cite{Dewitt73} by two different methods.

%

%

%

%
The Boltzmann expression for the chemical potential satisfies
the relation
\begin{eqnarray}
\exp (\mu_{i}^B/k_{\rm B}T) ={n_{i}\over \Omega_{i}}
\biggl({h^2N_0\over 2\pi k_{\rm B}TA}\biggr)^{3/2} \; .
\label{eq:varyag}
\end{eqnarray}


One has to introduce the corrections $\Delta_i$ in addition to
$\mu_{i}^{\rm B}$:
\begin{eqnarray}
\mu_{i} =Z\mu_{\rm p}^{\rm B}+(A-Z)\mu_{\rm n}^{\rm B}+
\Delta_{i}    \; ,
\label{eq:dalj}
\end{eqnarray}
where
\begin{eqnarray}
\Delta_{i}=\Delta_{i}^{\rm deg} +\Delta_{i}^{\rm Coul} +
\Delta_{i}^{\rm nuc} + \Delta_{i}^{\rm sur}\; .
\label{eq:kapor}
\end{eqnarray}
The corrections for degeneracy are:
\begin{eqnarray}
\Delta_{i}^{\rm deg}=Z(\mu_{\rm p}^0-\mu_{\rm p}^{\rm B}) +
(A-Z)(\mu_{\rm n}^0-\mu_{\rm n}^{\rm B}) \; ,
\label{eq:platok}
\end{eqnarray}
and for the Coulomb part:
\begin{eqnarray}
\Delta_{i}^{\rm Coul}=ZQ_p-Q_i \; .
\label{eq:shlyapa}
\end{eqnarray}

%
%
%

For nuclear interactions we have:
\begin{eqnarray}
\Delta_{i}^{\rm nuc} = &
 -B_{i}^0 +AW +(X_{\rm p}+X_{\rm n})
\biggl[Z{\partial W\over \partial X_{\rm p}}+(A-Z)
{\partial W\over \partial X_{\rm n}}\biggr] \hspace{40mm} \nonumber \\
 &
 + \biggl[-\chi_{i} +Z \sum _{A',Z'>1} {X_{A',Z'}\over
A'}\cdot {\partial \chi_{A',Z'} \over \partial X_{\rm p}}
+(A-Z) \sum_{A',Z'>1}{X_{A',Z'}\over A'}\cdot
{\partial \chi_{A',Z'} \over\partial X_{\rm n}}\biggr] 
\label{eq:uznik}
\end{eqnarray}

\citet{maz79} have argued that these expressions can often be
simplified. We take relation Eq.~(\ref{eq:uznik}) in full which requires an
additional loop of NSE iterations.
%

We assume the following correction to the surface energy:
\begin{eqnarray}
\Delta_i^{\rm sur} = W_{\rm sur}A^{2/3}
\biggl(1-{W\over W_{\rm nuc}}\biggr)^{1/2}\biggl[1-
(X_{\rm p}+X_{\rm n}){\rho \over \rho_{\rm nuc}}\biggr]^{4/3}
                                       \; .
\label{eq:lejka}
\end{eqnarray}
The expression actually used in the mass formula for the surface energy
may be much more complicated than just $W_{\rm sur}A^{2/3}$, but
since the term in braces tends to zero for vanishing $\rho$ this
correction does not influence the vacuum values of nuclear masses.


\subsection{Spin of ground state}\label{sec:spin}

    To find the spin of ground states of exotic and experimentally
unknown nuclei we have used the simple shell model with the Woods-Saxon potential
for defining the scheme of levels and
nuclear systematics. It is well known from
experiments and calculations using up-to-date versions of shell model
with residual interaction that the spin of ground states of even-even nuclei
is equal to zero. The spin of nuclear ground states of odd--$A$ nuclei is
determined generally by the spin of the uncoupled nucleon
(exceptions: $ ^{19}{\rm F}_9, \; ^{23}{\rm Na}_{11},
             \; ^{55}{\rm Mn}_{25}$).
That is why in the present paper we defined the ground state
spin of odd--$A$ nuclei by the spin of an uncoupled proton or neutron
according to experimental data and shell model calculations with
phenomenological coupling.

    The spin of ground state of odd-odd nuclei is defined by the spins of
the uncoupled proton,  $ j_p $, and neutron, $ j_n $, and
could not be calculated precisely
even on the base of the microscopic models of atomic nuclei.
%
%
%
%
That is why we have used, in the odd-odd case,  the simple
phenomenological Nordheim rules \citep{nor50}.

In the case of opposite spins of proton and neutron
($ j_n = l_n \pm 1/2, \;  j_p = l_p \mp 1/2$)
the spin of an odd-odd nucleus is equal to
\begin{eqnarray}
I =  j_n - j_p \; .
\label{eq:nechet}
\end{eqnarray}
In the other case ($ j_n = l_n \pm 1/2, \; j_p = l_p \pm 1/2 $) it is rather
difficult to find the right value of ground state spin because it can fall
into the wide range:
\begin{eqnarray}
j_n - j_p \leq I \leq j_n + j_p \;.
\label{eq:nchparl}
\end{eqnarray}
But it is well known from systematics that $ I \approx j_n + j_p $ mainly. For
simplicity we have chosen $ I = j_p + j_n $.

    The value of the ground state spin should not strongly influence
the results of NSE calculations because in our model the nuclei are mostly
in the excited states.
We have done a calculation
using another approach of the ground state
spin calculation \citep{eng91} to check the influence.
It is also based on the simple shell model
level scheme, and significantly differs from ours only in the case of
odd-odd nuclei, for which the spin is defined as: $ I=(j_p + j_n + 1)/2$.

We find
that the calculated results depend weakly on the different approaches
to ground state spin calculation
only in the region with the significant amount of "heavy" nuclei,
where the role of
odd-odd nuclei rises. Since the most abundant nuclei
are even-even and odd--$A$ ones under NSE conditions, one can use any simple
estimation of  the ground state spin of nuclei under these conditions.
For $T_9 < 10$ and low $\rho$ one should prefer our model of
ground state spin evaluation, or an ``exact'' calculations of nuclear
characteristics from first principles.

As for the role of the excited states, we agree with the conclusions of \citet{Liu07}
that the effect is not very large.
They used a reliable method of  nuclear
partition function calculation, based on the Fermi gas formula.
However, the question deserves further investigation,
especially with account of new data from \citet{RTK97,RT2000,Rauscher2003}.
The used in \citet{Liu07} is related to an energy-dependent level density parameter with microscopic
correction to a nuclear mass model \citep{RTK97,RT2000,Rauscher2003}.
Our code may incorporate various options for the partition functions.





\subsection{Nuclear masses}\label{sec:mass}

We adopt the table of nuclear masses (KTUY) by \citet{kou05}
for our EOS calculations.
The prediction of nuclear masses by \citet{kou05} is based on
the mass formula composed of macroscopic and microscopic terms \citep{kou00},
which treat deformation, shell and even-odd energies.
It covers $\sim$9000 species of nuclei (Z $\ge$ 2 and N $\ge$ 2)
in the nuclear chart.
We use the extended mass table containing $\sim$20000 species \citep{kou07}
to cover the neutron- and proton-rich regions in the nuclear
statistical equilibrium.
The root-mean-square deviation from experimental masses is 666.7 keV.
We examine the dependence on the nuclear mass by adopting other tables of
the nuclear masses by \citet{HilfGT76}.
The mass data by \citet{mol95} will be implemented in the code.
Those mass tables predict the experimental masses very well
around the stability line in the nuclear chart, however,
they provide different predictions in neutron-rich region
away from the stability.
It would be interesting to examine whether these differences
appear in the composition of dense matter in supernova core.

\subsection{Shen equation of state}\label{sec:shen}

We adopt the table of EOS by \citet{she98a,she98b}
to investigate the influence of mixture of nuclei
on the equation of state.
The Shen EOS is a set of thermodynamical quantities of dense matter
under the wide range of density, proton fraction and
temperature for supernovae.
It has been widely used for numerical simulations
of core-collapse supernovae \citep{sum05,sum06,jan05,bur06}
and other astrophysical phenomena \citep{ros03}.
The equation of state is calculated by the relativistic mean field
theory \citep{ser86}, which is based on the relativistic Br{\"u}ckener Hartree
Fock theory \citep{bro90} and is constrained by the experimental data of
neutron-rich nuclei \citep{sug94}.
The nuclei in dense matter is described within the local density
approximation assuming one species of nucleus surrounded by
neutrons, protons and alpha particles in the Wigner-Seitz cell.
The basic behaviour of the Shen EOS in supernova core as well as
the comparison with the another set of EOS by \citet{lat91}
can be found in \citet{sum04,sum05}.

\subsection{Beta-equilibrium}\label{sec:beta}

Although we fix the value of $Y_{\rm e}$ in the current calculations,
we can impose the condition of beta equilibrium in the Saha approach.
We add the formulation here for the applications to proto-neutron stars.

The condition of beta-equilibrium is governed by the relation
\begin{eqnarray}
  \mu_{\rm p} + \mu_{\rm e} = \mu_{\rm n} + \mu_\nu \; .
\label{milka}
\end{eqnarray}
Therefore, one has to calculate chemical potentials of neutrinos
and electrons in an additional loop of iterations.

For calculating lepton chemical potentials one can use simple
relations derived by
\citet{nad74a,nad74b} \citep[see also][]{bli87,bli88}:
\begin{eqnarray}
n-\tilde n ={\partial P\over \partial \mu}
\left\vert \matrix{& \cr
                   T \cr} \right.
                   ={g\over 6\pi^2} \Bigl(\mu^3+\pi^2 \mu T\Bigr)   \; .
\label{zhaba}
\end{eqnarray}
Here $n$ is a fermion number in the unit volume, i.e., the fermion
concentration, $\tilde n$ is an anti-fermion concentration.
$n-\tilde n$ is a charge of the unit volume, for example, for the
electrons and it is a lepton charge for neutrinos ($g=2$ for electrons
and $g=1$ for neutrinos). It is possible to
express the lepton chemical potentials $\mu$ through the radicals
if $n-\tilde n$ and $T$ are given.
In actual calculations it is faster to do this by Newton iterations.

\section{Results}\label{sec:result}

We report the numerical results in a set of selected conditions
for density, $\rho$, electron fraction, $Y_{\rm e}$, and temperature, $T$.
We show the properties of dense matter at typical conditions
($\rho$=10$^{10}$--10$^{13}$ g/cm$^3$) during the core collapse,
where the composition of nuclei and free protons are crucial
to determine the electron capture.
We remark that the present framework is generally applicable
in the wide range of conditions, which are necessary for core-collapse
supernova simulations.
However, it is limited up to $Y_{\rm e}$ $\sim$0.3
due to the coverage of nuclear mass models at neutron-rich region.
The limited range of
low $T$ $\sim$2 MeV and high $\rho$ $\sim$10$^{13}$ g/cm$^3$
is due to the consideration of nuclear interaction and
the convergence of iterations.

We show, at first, the general behaviour of equation of state
in the current framework with the mass formula by \citet{kou05}
in \S \ref{sec:result-general}.
Next, we compare the equation of state obtained by the current Saha-approach
with that of Shen EOS \citep{she98b} in \S \ref{sec:result-compare}.
We examine the difference between the two nuclear mass models.
We also compare with the results calculated in the formulation of \citet{maz79}.
In \S \ref{sec:result-composition}, we show the composition of nuclei
in the nuclear chart in the situations of supernova core.

\subsection{General behaviour}\label{sec:result-general}

\begin{figure}
\includegraphics[width=0.45\linewidth]{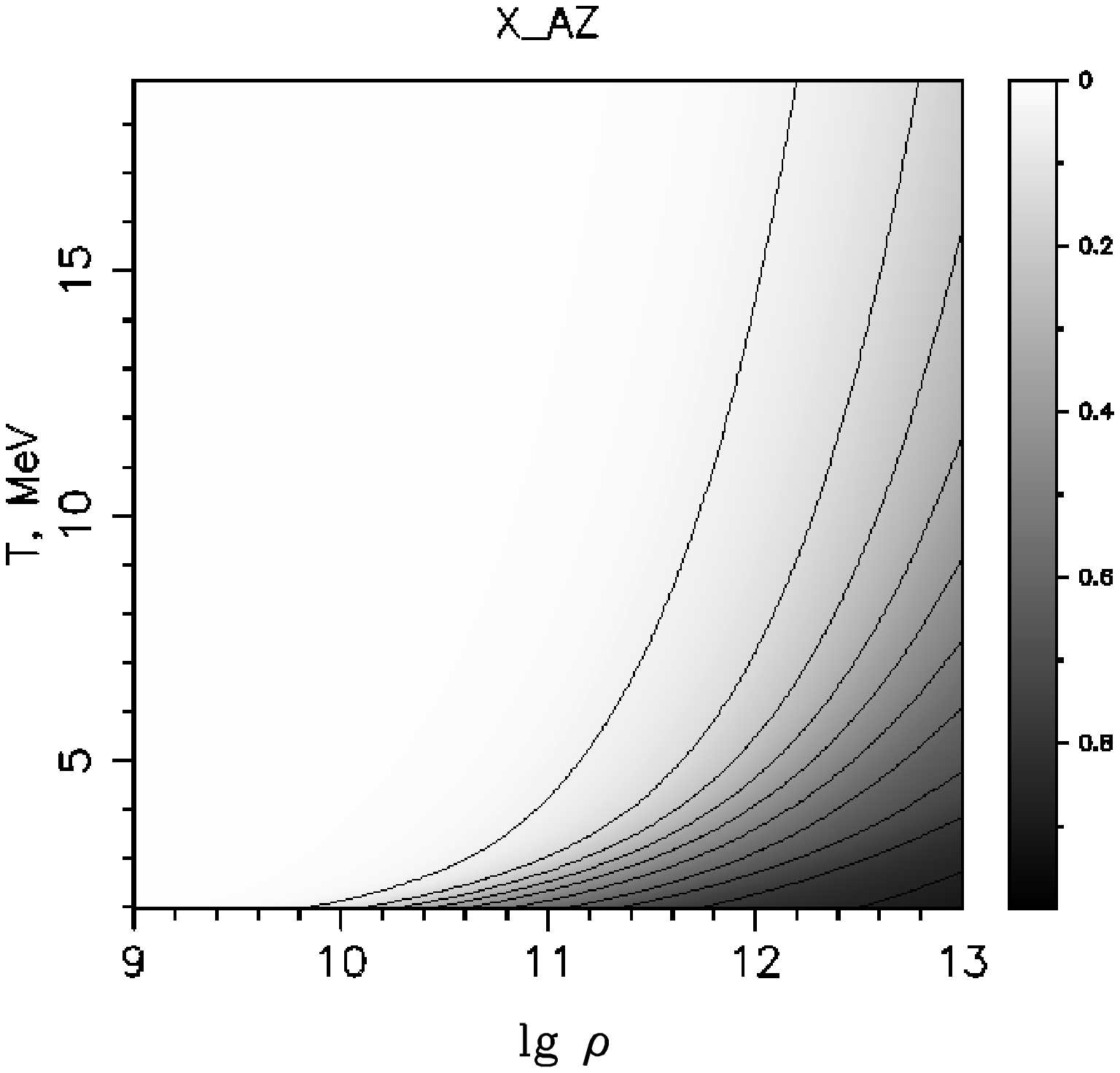}
\includegraphics[width=0.45\linewidth]{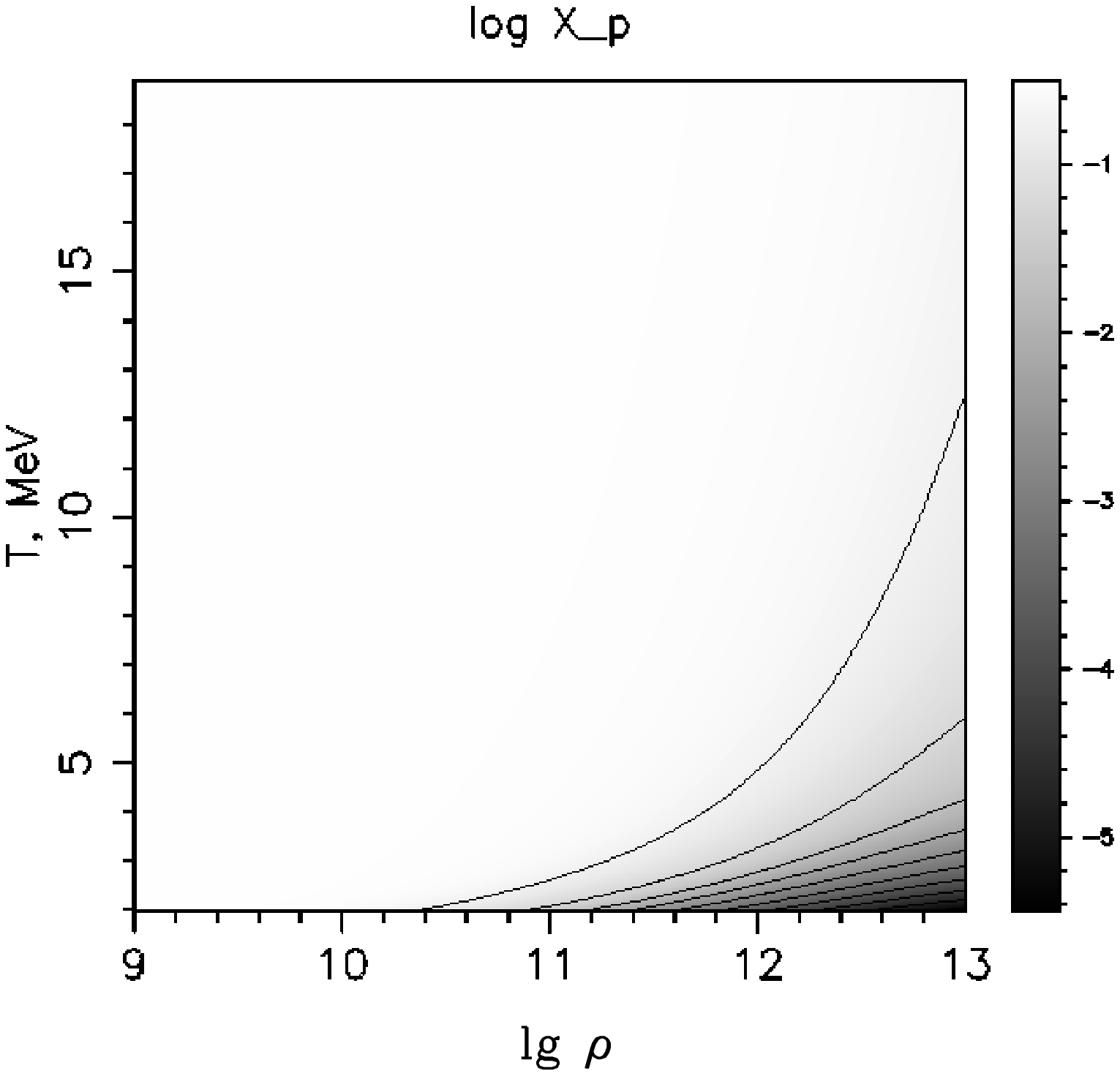}
\caption{\noindent  The isocontours of
the mass fraction of heavy nuclei, $X_{AZ}$, (left) and
free protons, $X_p$, (right)
for $Y_{\rm e}=0.316$ are shown in the plane of density and temperature.
        }
\label{fig:2d-xaz-y316}
\end{figure}

\begin{figure}
\includegraphics[width=0.45\linewidth]{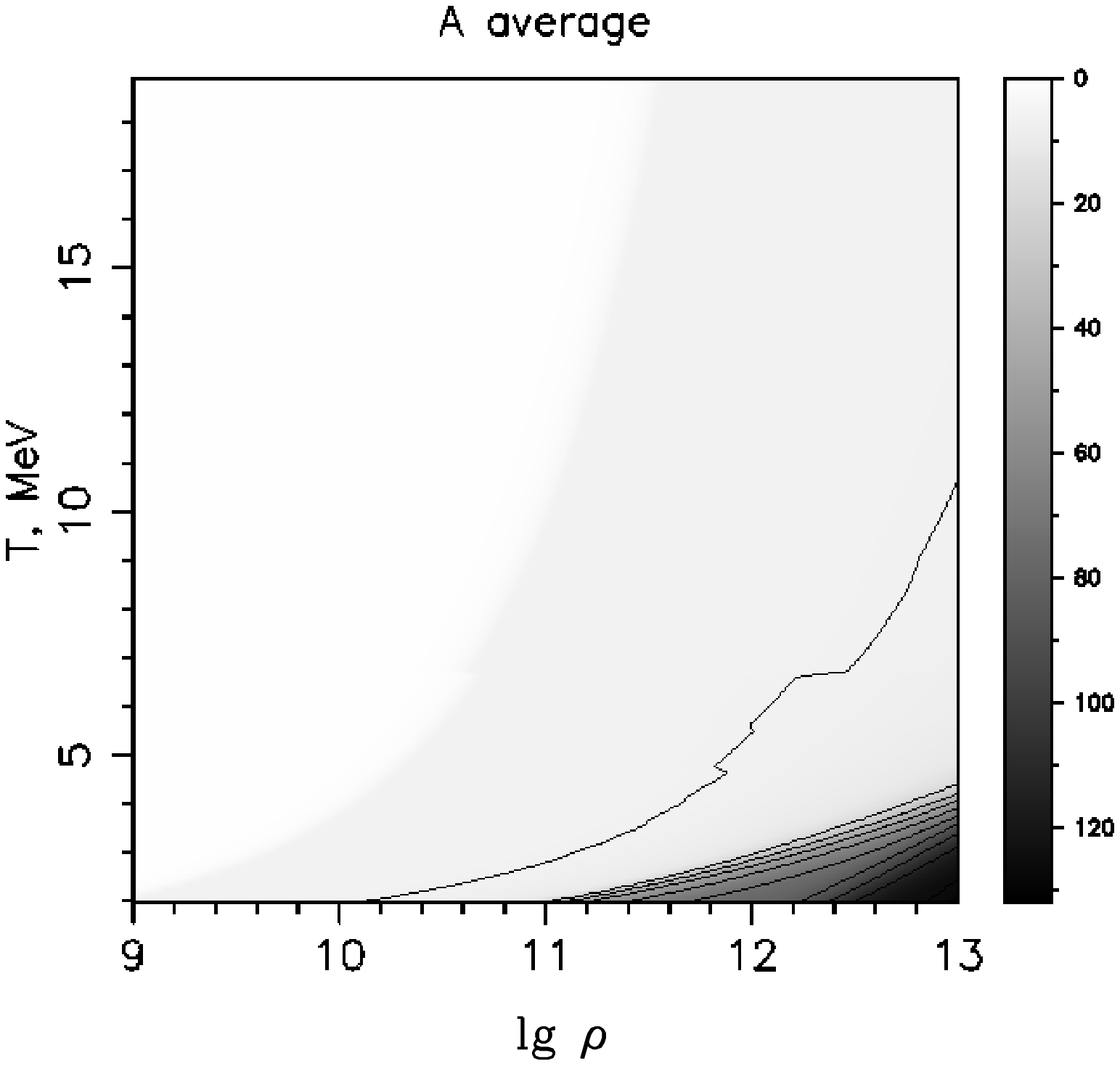}
\includegraphics[width=0.41\linewidth]{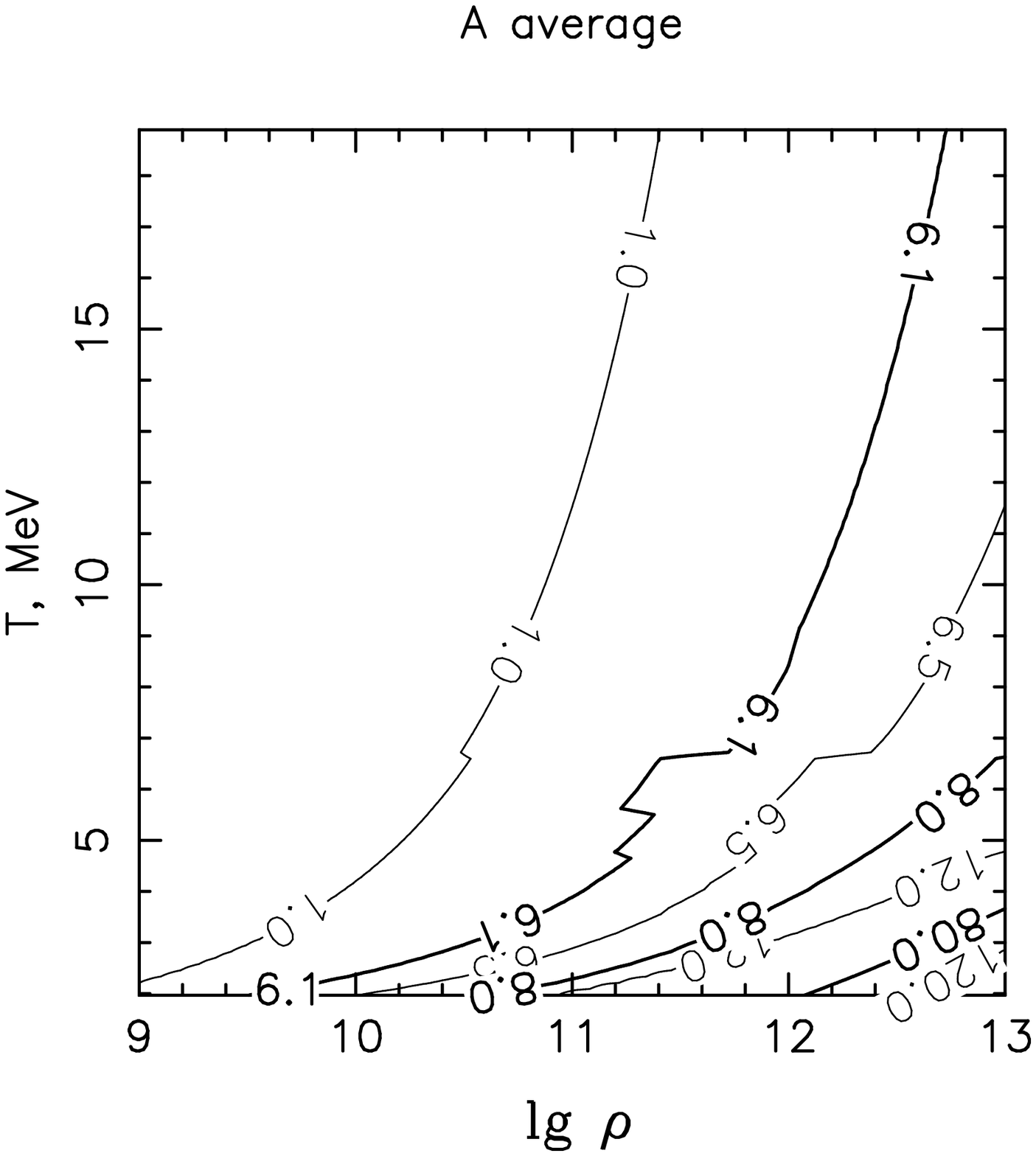}
\caption{\noindent  The isocontours of
the average mass number, $A$, of heavy nuclei,
for $Y_{\rm e}=0.316$.
        }
\label{fig:2d-a-y316}
\end{figure}

\begin{figure}
\includegraphics[width=0.45\linewidth]{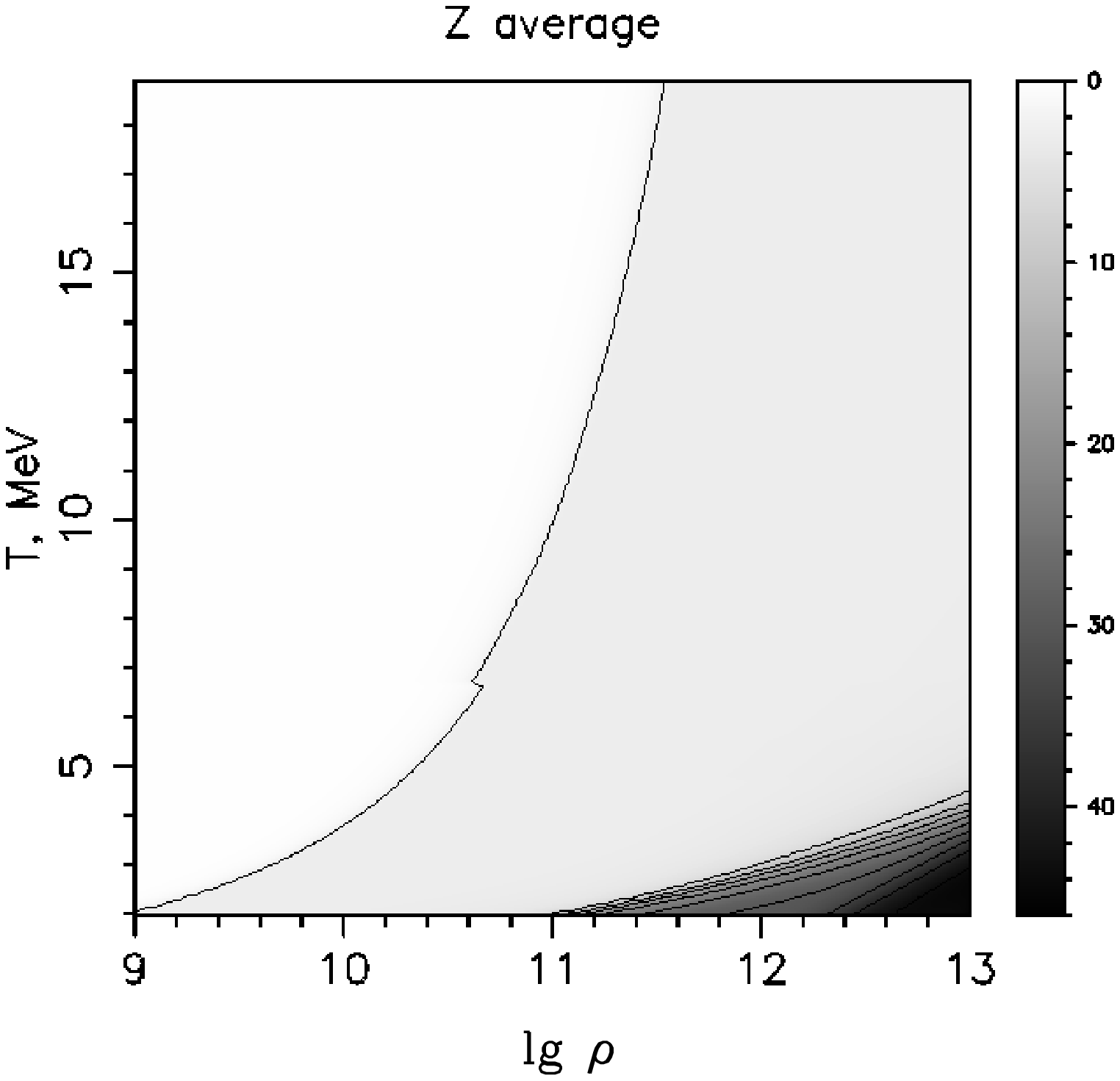}
\includegraphics[width=0.41\linewidth]{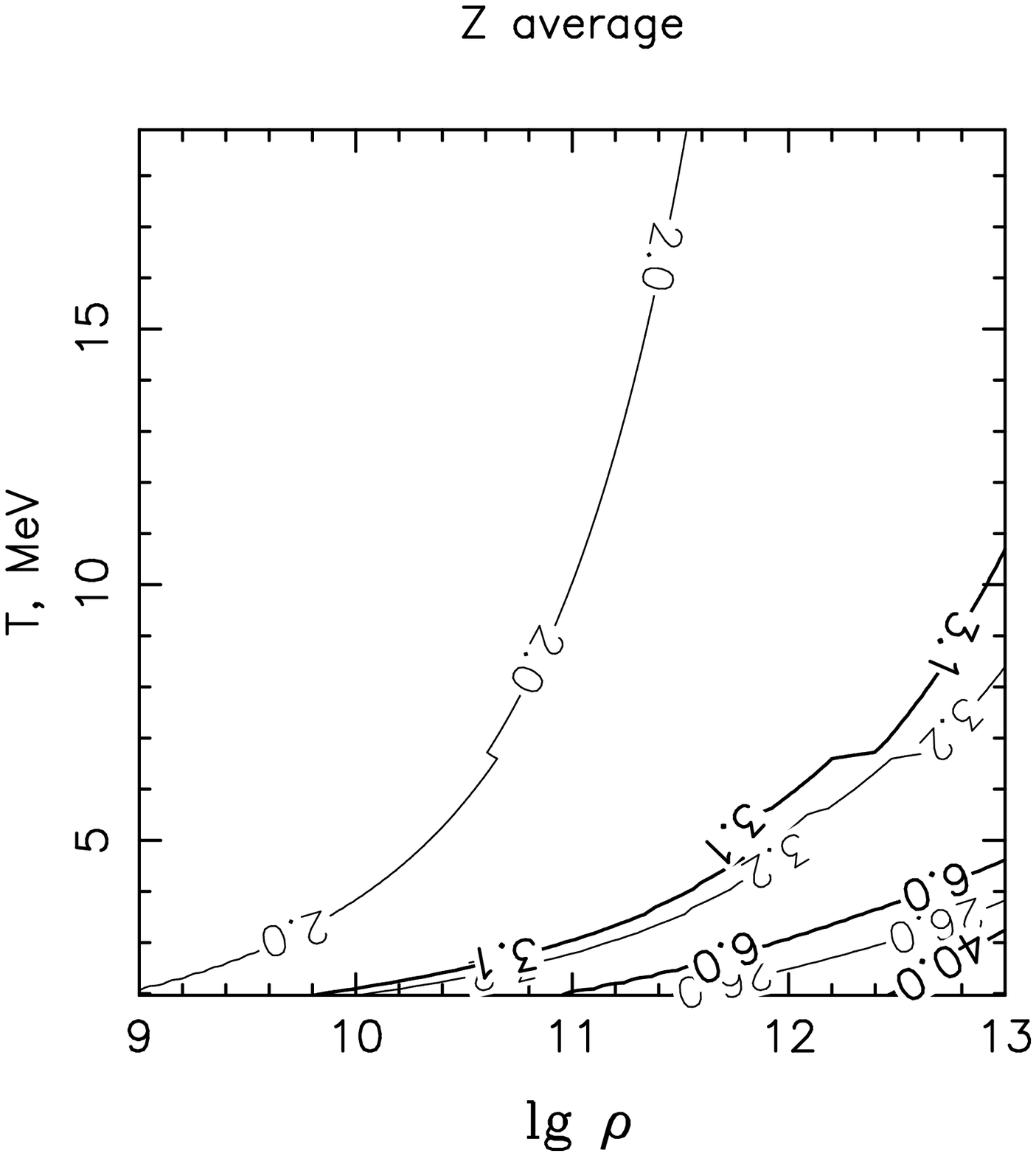}
\caption{\noindent  The isocontours of
the average proton number, $Z$,  of heavy nuclei,
for $Y_{\rm e}=0.316$.
        }
\label{fig:2d-z-y316}
\end{figure}

\begin{figure}
\includegraphics[width=0.45\linewidth]{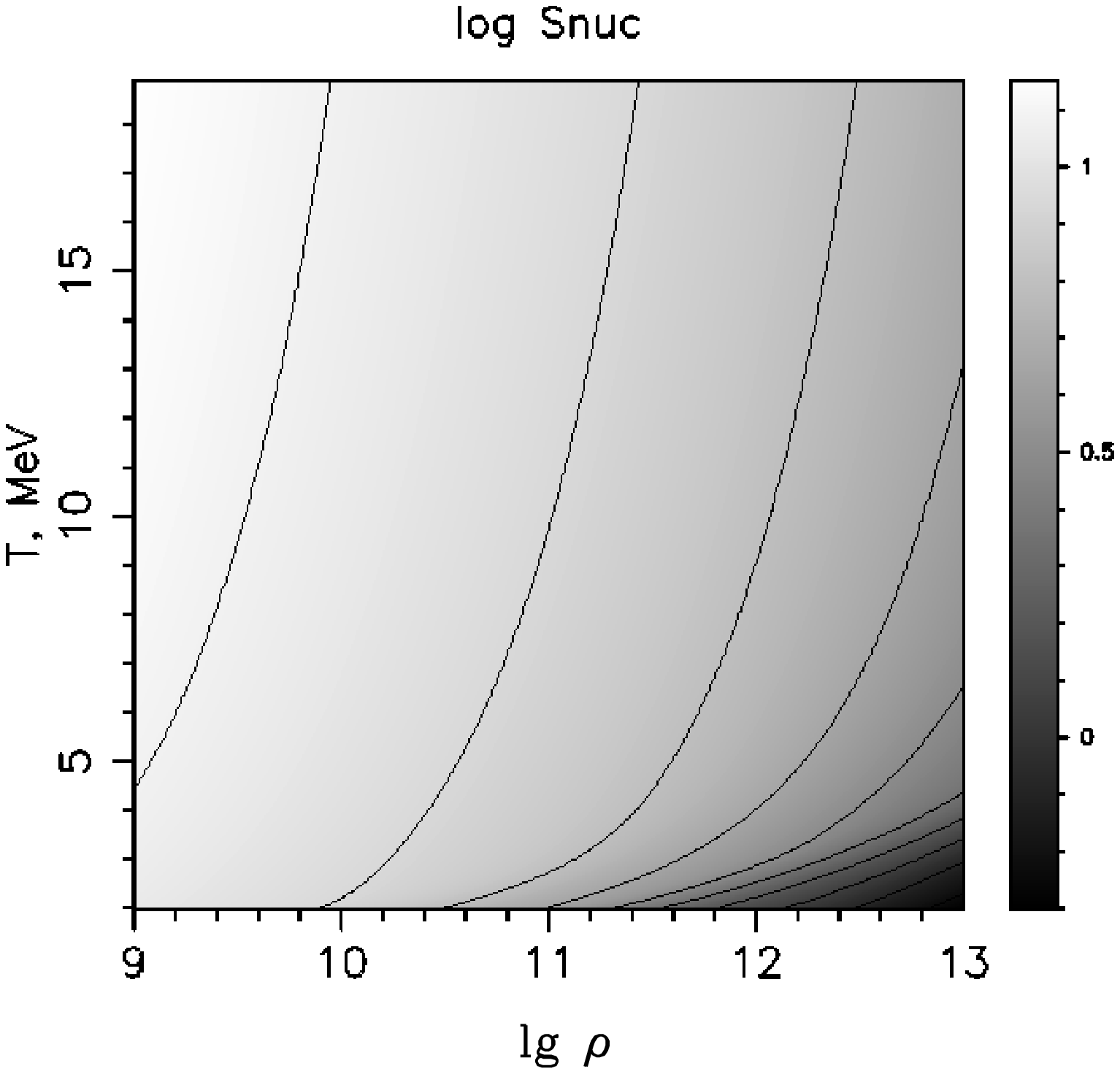}
\includegraphics[width=0.42\linewidth]{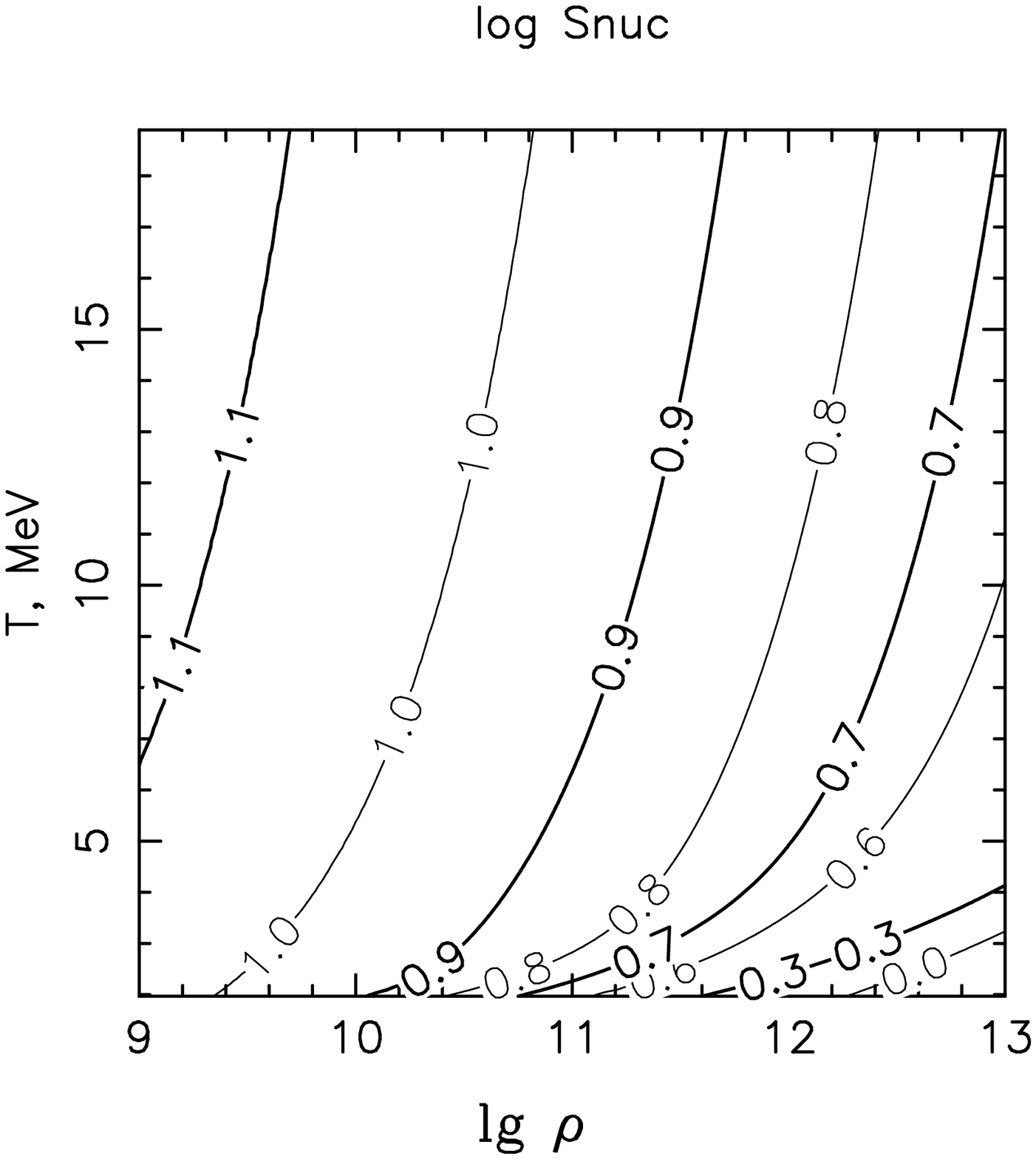}
\caption{\noindent  The isocontours of
the nuclear contribution to entropy, $S_{\rm nuc}$,
for $Y_{\rm e}=0.316$ in the plane of density and temperature.
        }
\label{fig:2d-sn-y316}
\end{figure}

We show, in Figs. \ref{fig:2d-xaz-y316}, \ref{fig:2d-a-y316}, \ref{fig:2d-z-y316} and \ref{fig:2d-sn-y316},
the global features of dense matter at $Y_{\rm e}=0.316$
for a wide range of density and temperature relevant to collapsing supernova cores.
We have calculated also the cases with $Y_{\rm e}=0.398$ and $Y_{\rm e}=0.473$,
which are not shown here.
The values of $Y_{\rm e}$ are selected so as to match with the values
in the original table of Shen EOS.
We see very smooth variations of calculated quantities
in the range of density and temperature shown in the plot.
There are some wiggles in the plots of the average of mass number, $A$, and
proton number, $Z$ in Figs. \ref{fig:2d-a-y316}, \ref{fig:2d-z-y316} due to the shell effects.
One can recognise the appearance of nuclei at low density $\sim$ 10$^{10}$ g/cm$^{3}$
at the lowest temperature.
This is different from the Shen EOS as we will see in the next subsection.

\subsection{Comparisons among models}\label{sec:result-compare}

We compare the current results with those in the Shen EOS
to see the effect of multi-composition with respect to
the one-species treatment, which has been used
in most of the supernova EOS's.
We examine also the dependence on the nuclear mass models
by comparing the cases by \citet{kou05} and \citet{HilfGT76}.
We set here $Y_{\rm e}=0.316$ and $T$=2 MeV
which roughly corresponds to the neutron-rich supernova core.
This $T$ value also corresponds to the lower $T$ boundary in the figures
shown above.

\begin{figure}
\includegraphics[width=0.45\linewidth]{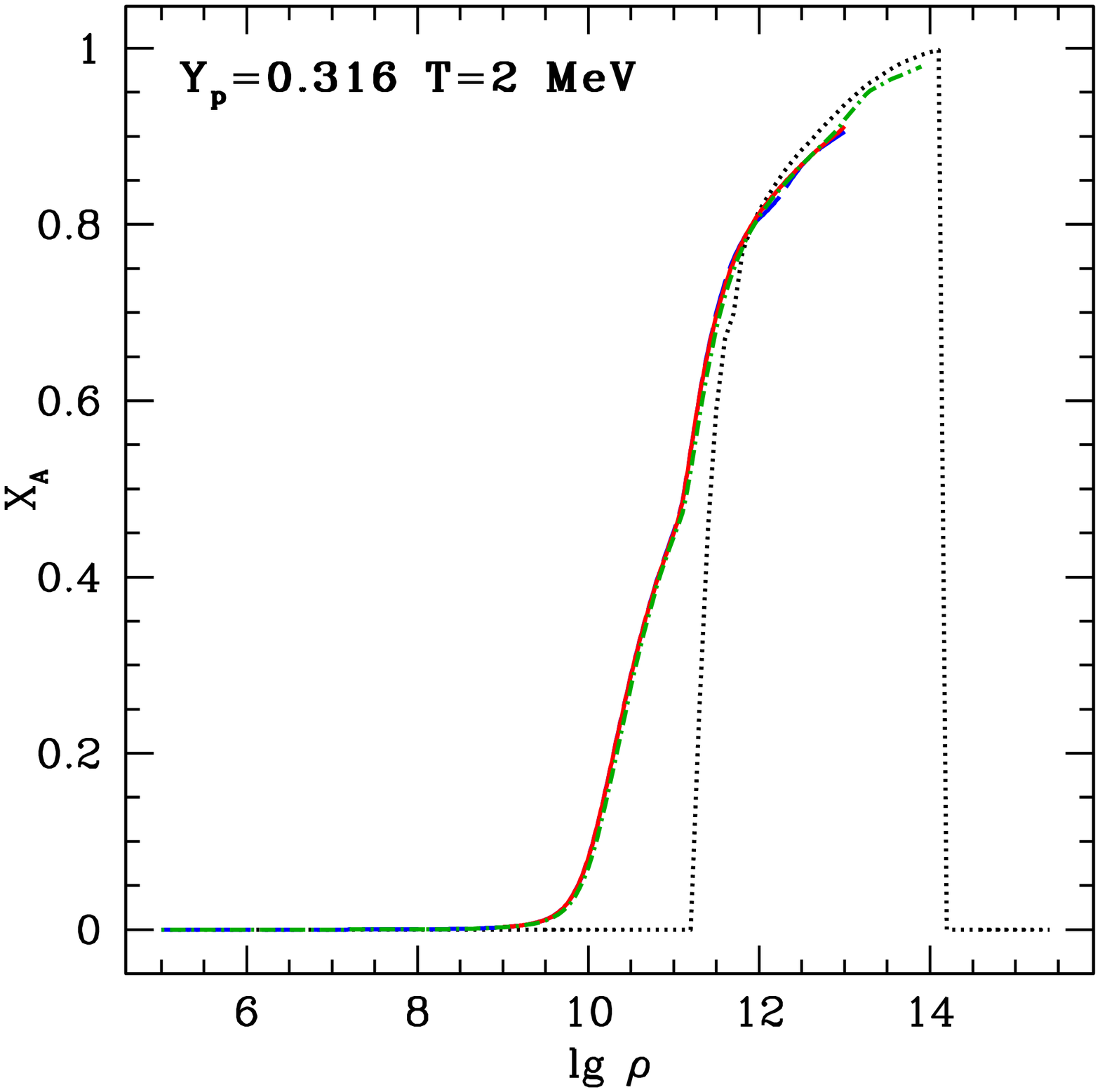}
\includegraphics[width=0.45\linewidth]{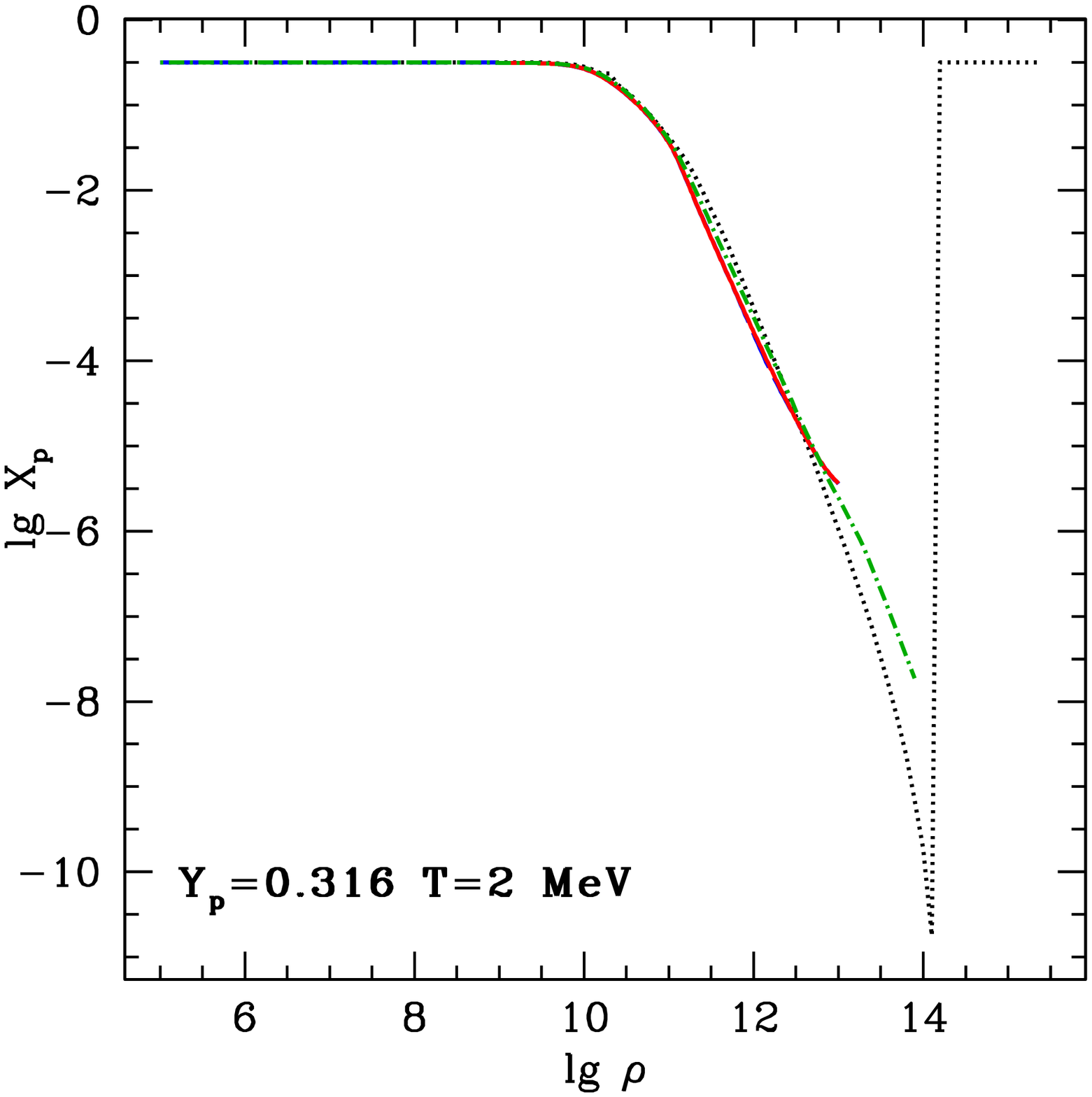}
\caption{\noindent  The mass fraction of nuclei, $X_A$, (left)
and free protons, $X_p$, (right)
as a function of density
for $Y_{\rm e}=0.316$ and $T=2$ MeV.
The solid and dashed lines denote the NSE calculation with mass-formulae by \citet{kou07}
and \citet{HilfGT76}, respectively.
The dash-dot lines are obtained using \citet{kou07}, but with a simplified treatment of relation  Eq.~(\ref{eq:uznik})
as has been done by \citet{maz79}.
The dotted lines are the results from \citet{she98b}.
        }
\label{fig:1d-xap-y316t2}
\end{figure}

\begin{figure}
\centering
\includegraphics[width=0.45\linewidth]{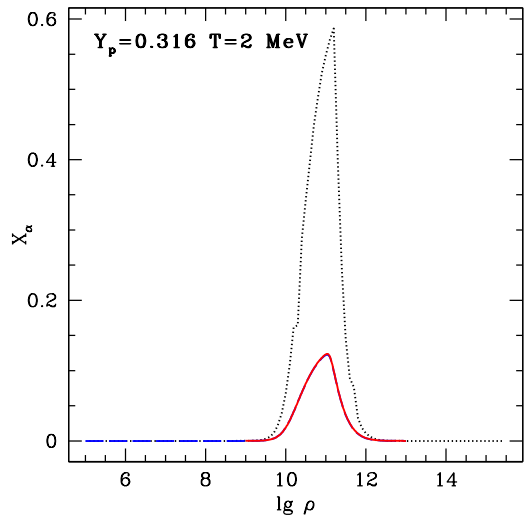}
\caption{\noindent  The mass fraction of alpha-particles, $X_\alpha$, as a function of density
for $Y_{\rm e}=0.316$ and $T=2$ MeV.
The notation is the same as in Fig.~\ref{fig:1d-xap-y316t2}.
        }
\label{fig:1d-xal-y316t2}
\end{figure}

\begin{figure}
\includegraphics[width=0.45\linewidth]{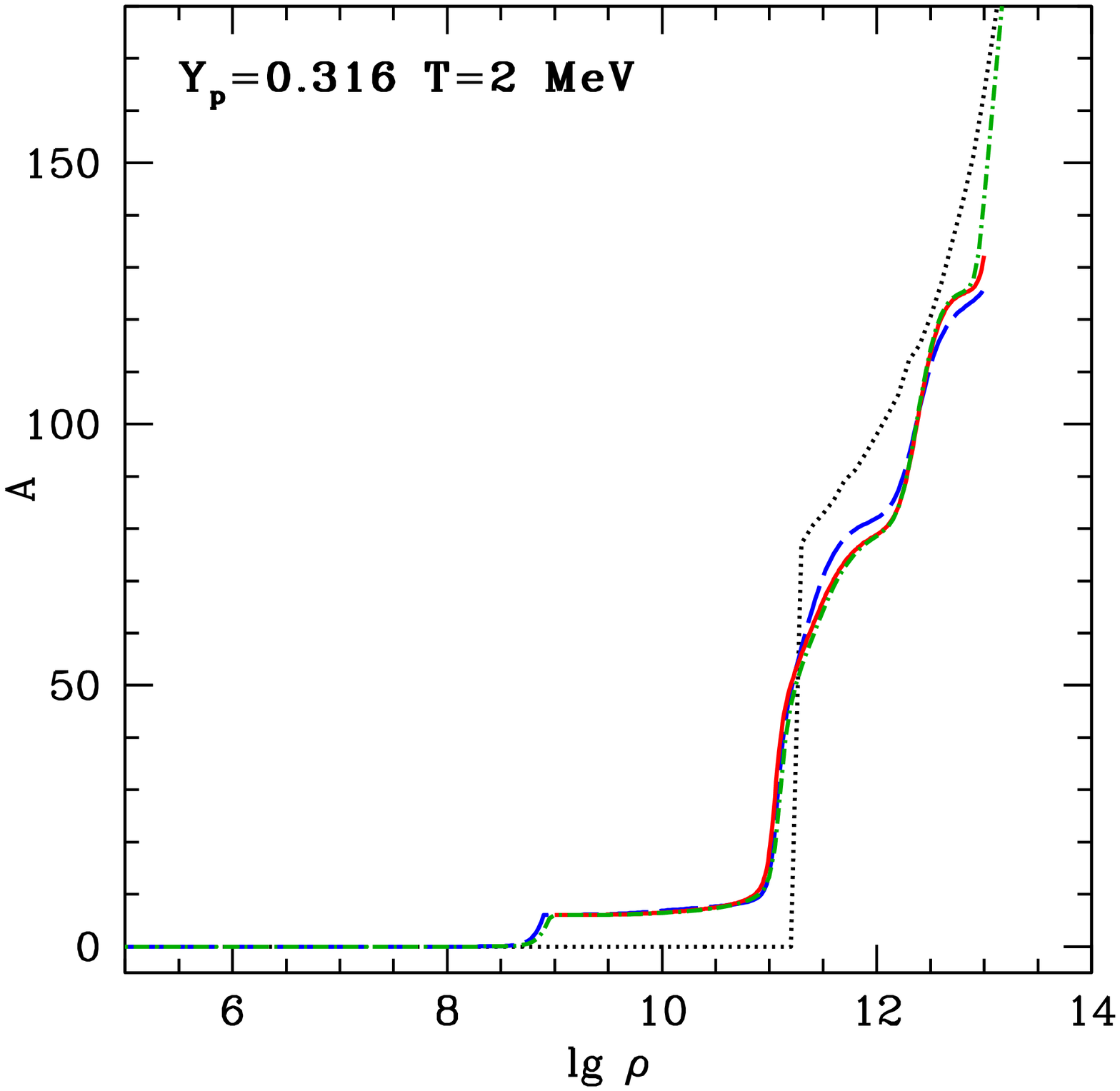}
\includegraphics[width=0.45\linewidth]{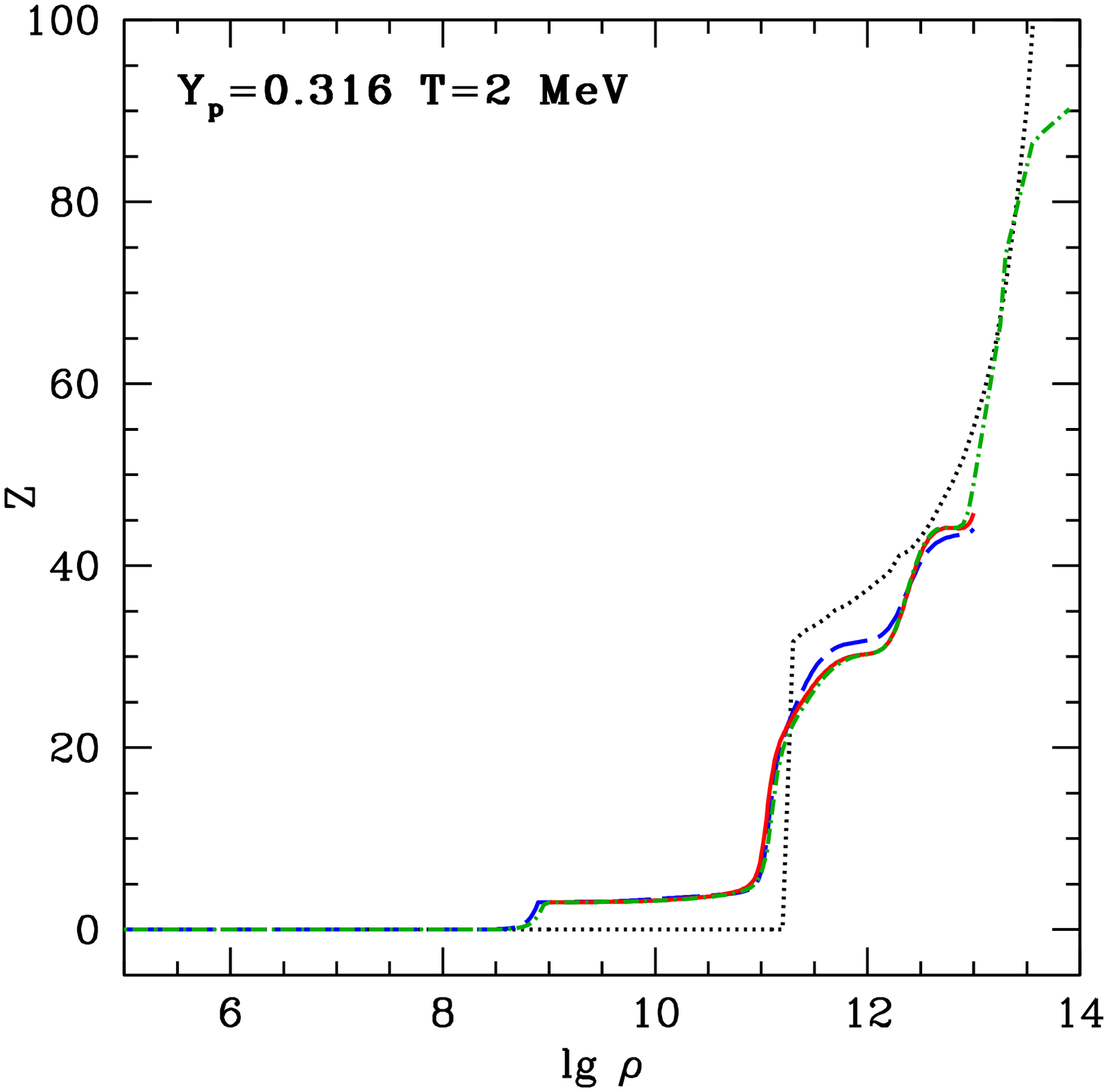}
\caption{\noindent  The average mass number, $A$, (left)
and average proton number, $Z$, (right)
of nuclei as a function of density
for $Y_{\rm e}=0.316$ and $T=2$ MeV.
The notation is the same as in Fig.~\ref{fig:1d-xap-y316t2}.
        }
\label{fig:1d-az-y316t2}
\end{figure}

We show that the mass fractions of nuclei, free protons and
alpha particles as a function of density
in Figs. \ref{fig:1d-xap-y316t2} and \ref{fig:1d-xal-y316t2}.
As the density goes higher, we have found the nuclei appear
earlier than the Shen EOS does.  
This is because the nuclei other than alpha particles
appear abundantly in the treatment of multi-composition.
In the Shen EOS, on the other hand, alpha-particle appear
more in this density region $\sim$10$^{11}$ g/cm$^{3}$
as seen in Fig. \ref{fig:1d-xal-y316t2}.
Figure \ref{fig:1d-az-y316t2} elucidates clearly this
difference further.
From $\sim$10$^{9}$ g/cm$^{3}$, the Saha-treatment
provides nuclei of average mass number $A \sim 10$.
At densities higher than $\sim 3 \times $10$^{11}$ g/cm$^{3}$,
the average mass number is smaller than that of representative
nuclei in the Shen EOS.
The one-species treatment may overestimate the mass number
in this respect.
We note that the average mass number in the Saha-treatment
does not include the contribution of alpha particles,
which are separately treated in the Shen EOS.

\begin{figure}
\includegraphics[width=0.45\linewidth]{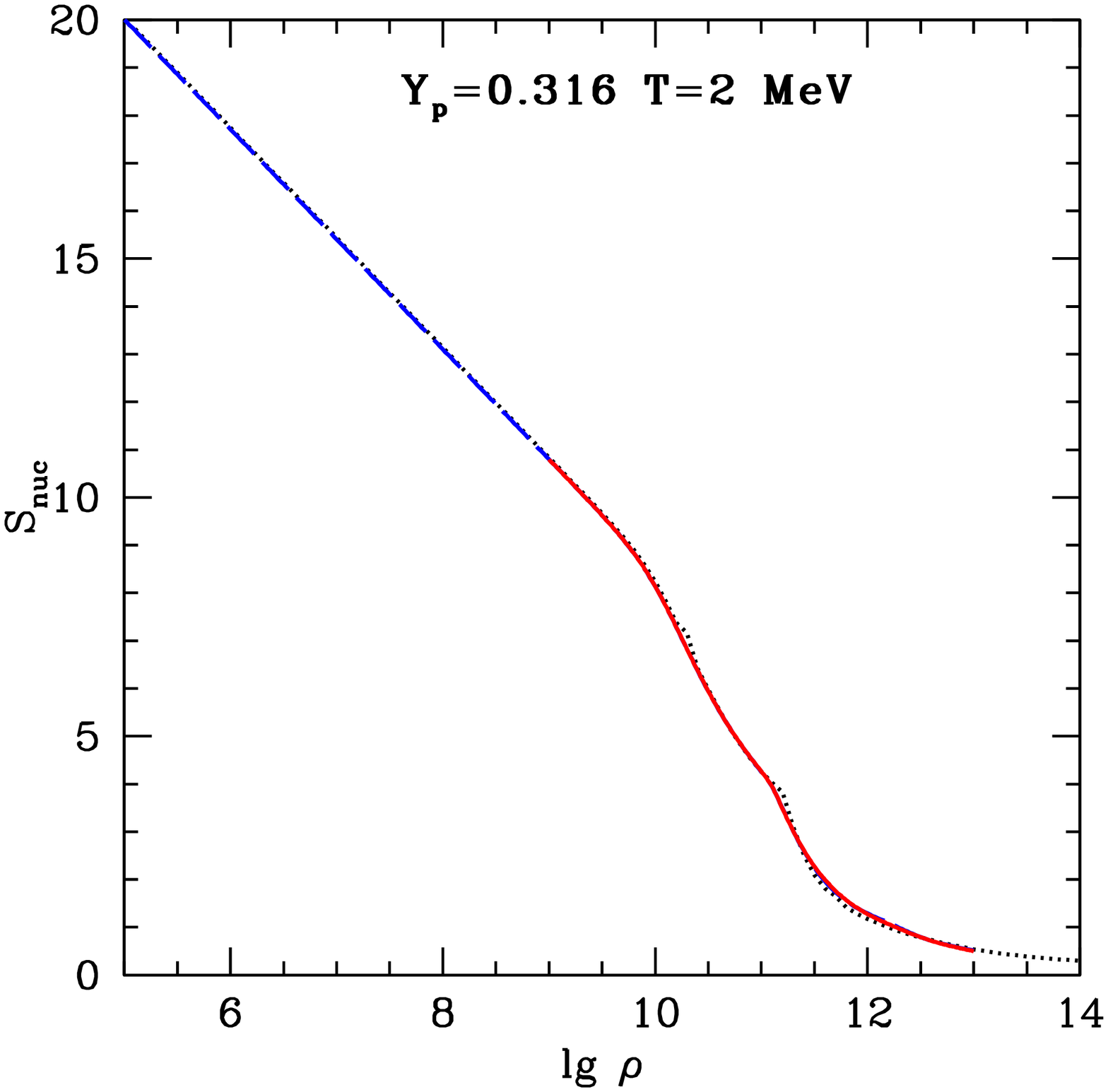}
\includegraphics[width=0.45\linewidth]{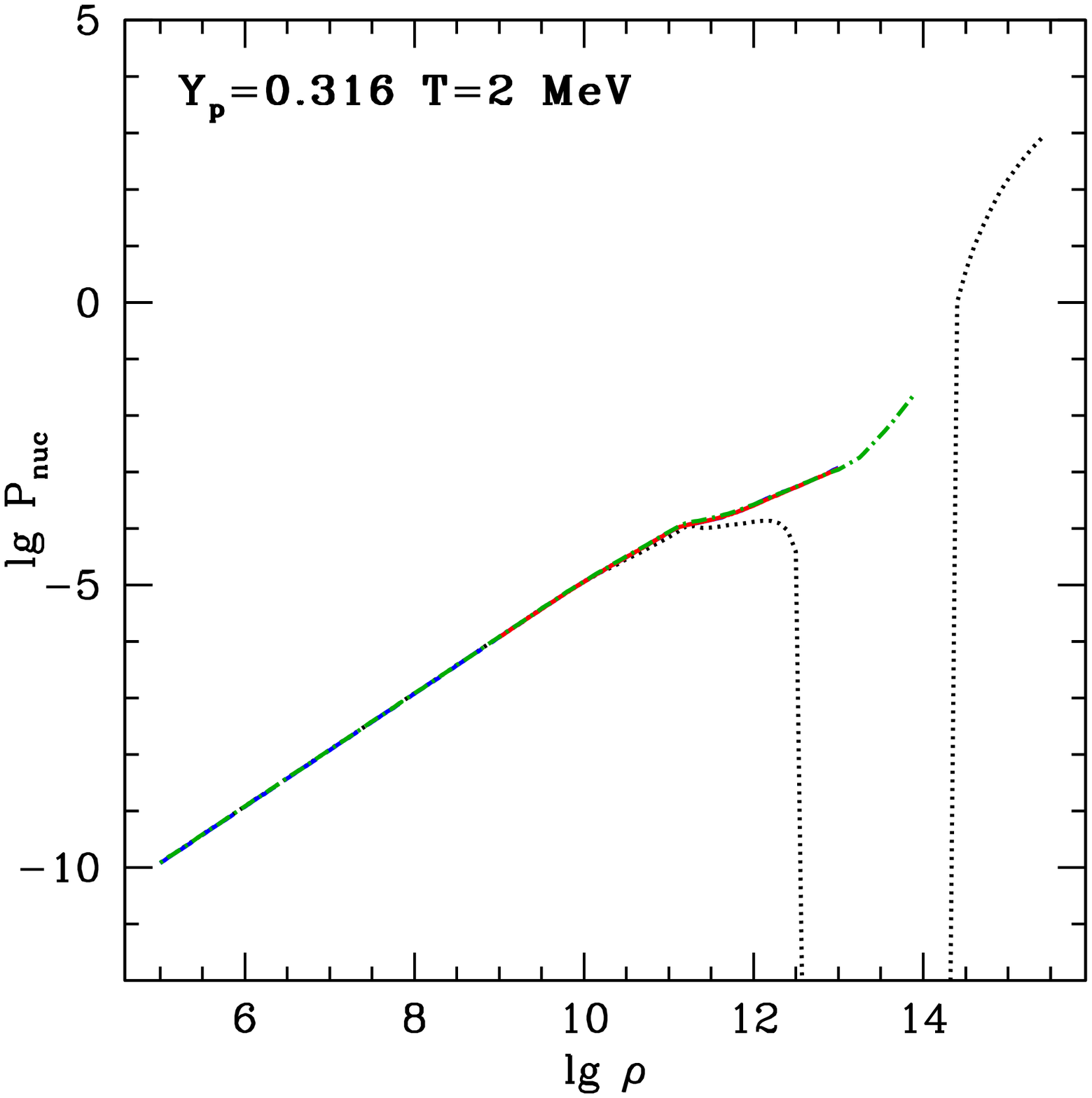}
\caption{\noindent  The  entropy, $S_{\rm nuc}$ (left), and pressure $P_{\rm nuc}$ (right),
of nuclei as functions of density for $Y_{\rm e}=0.316$ and $T=2$ MeV.
The notation is the same as in Fig.~\ref{fig:1d-xap-y316t2}.
        }
\label{fig:1d-sn-y316t2}
\end{figure}

The thermodynamical quantities, i.e. entropy and pressure of
nuclear contribution, are shown in Fig. \ref{fig:1d-sn-y316t2}.
The entropy of the Shen EOS is in accord very well
with the current result.
The pressure of the Shen EOS has trough around 10$^{13}$ g/cm$^{3}$
because of the decrease of nuclear part of free energy.
As it has been already reported in \citet{she98a},
this is due to the increasing binding energy of nuclei.
We note that one should add lepton and radiation contributions
in addition, therefore, the total pressure is always positive.

\begin{figure}
\includegraphics[width=0.45\linewidth]{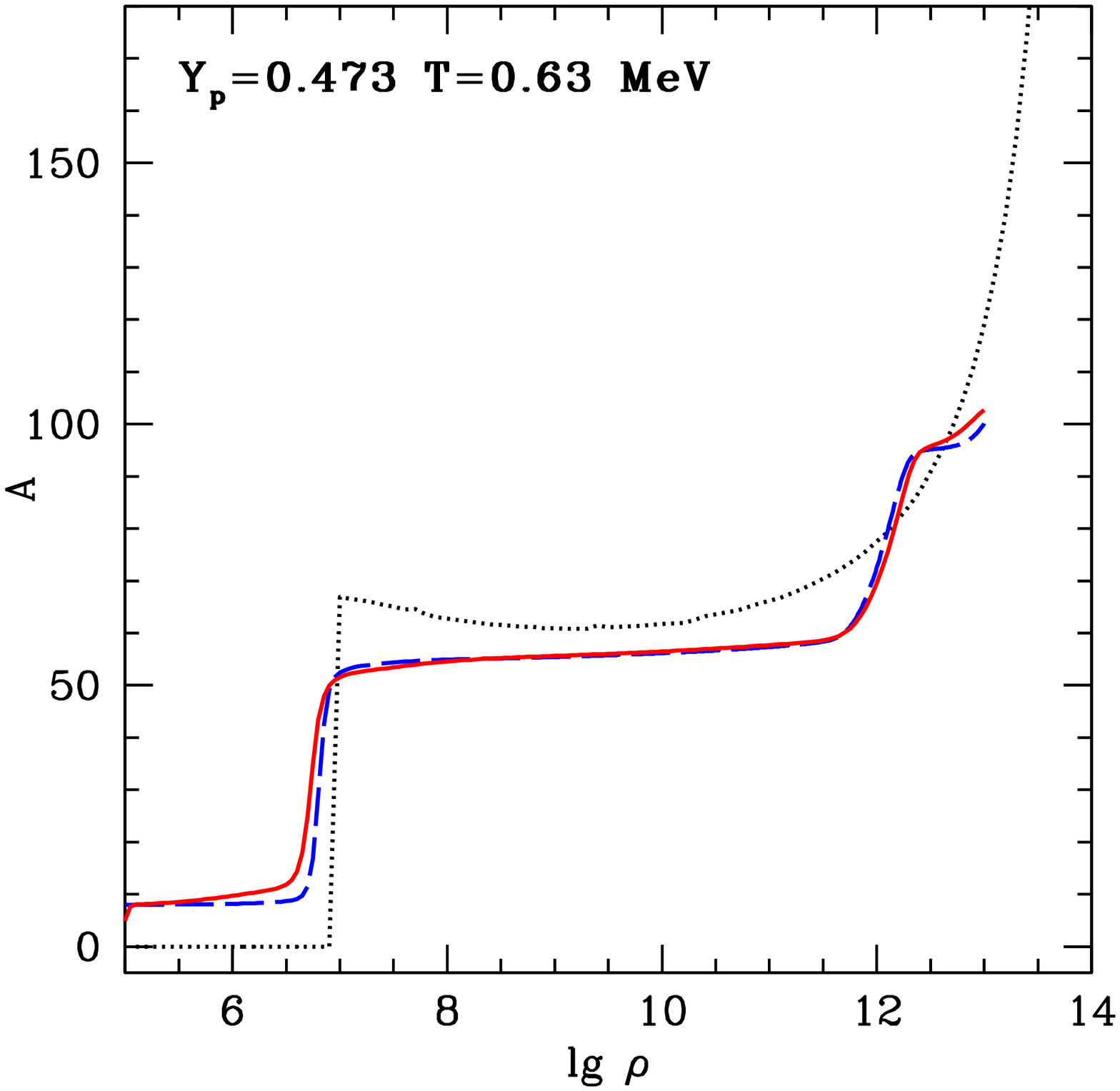}
\includegraphics[width=0.45\linewidth]{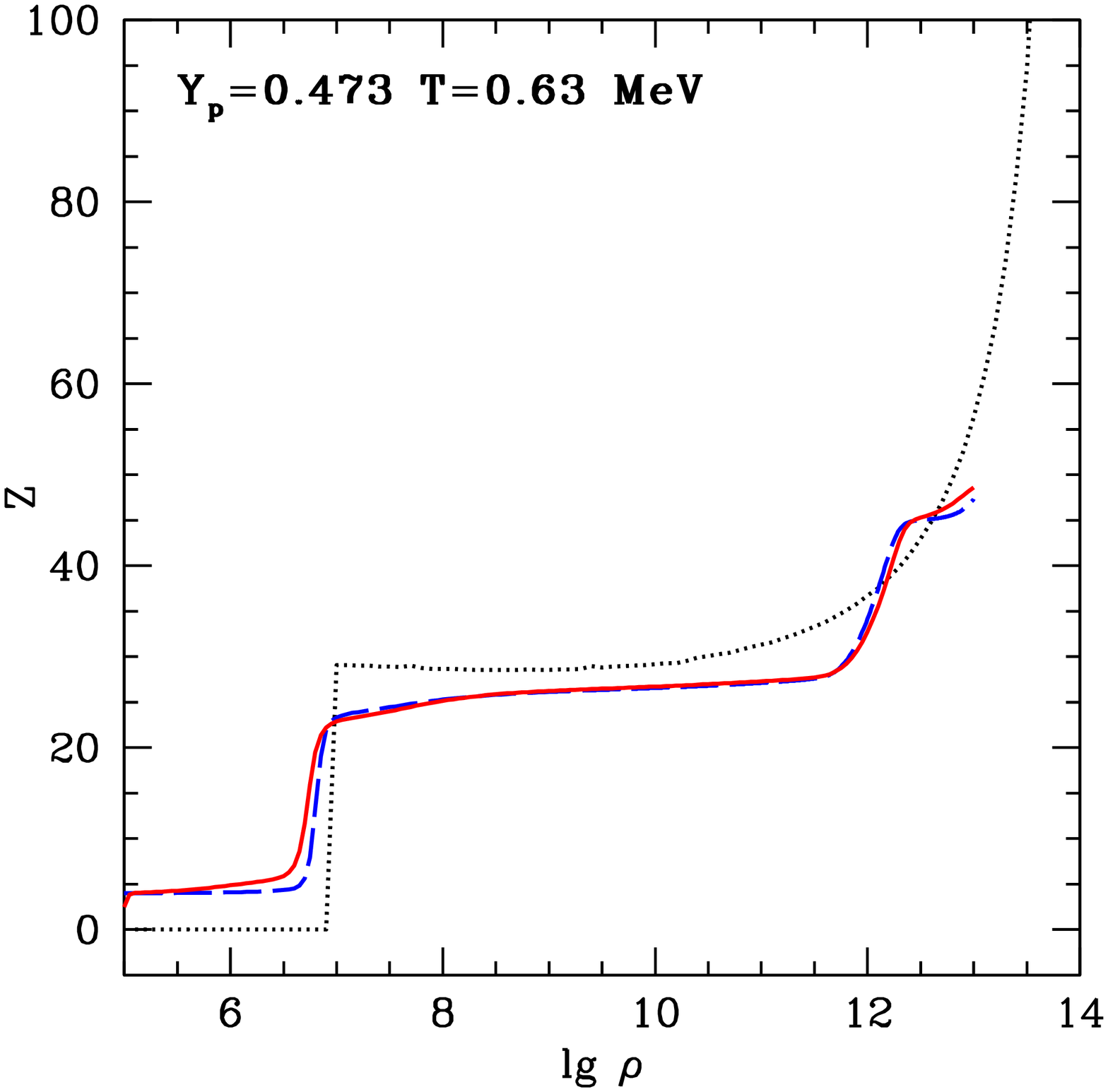}
\caption{\noindent  The average mass number, $A$, (left)
and average proton number, $Z$, (right)
of nuclei as a function of density
for $Y_{\rm e}=0.473$ and $T=0.63$ MeV.
The notation is the same as in Fig.~\ref{fig:1d-xap-y316t2}.
        }
\label{fig:1d-az-y473t063}
\end{figure}

In Fig. \ref{fig:1d-az-y473t063}, we show the average mass number
and proton number of nuclei for $Y_{\rm e}=0.473$ and $T$=0.63 MeV,
which is more closer to the condition at centre of the initial
progenitors.
We can see that mass number (proton number also) in the current treatment
is larger at low density and smaller at high density.
This is quite similar to the case of more neutron-rich and higher temperature,
$Y_{\rm e}=0.316$ and $T$=2 MeV in Fig. \ref{fig:1d-az-y316t2}.

\begin{figure}
\centering
\includegraphics[width=0.45\linewidth]{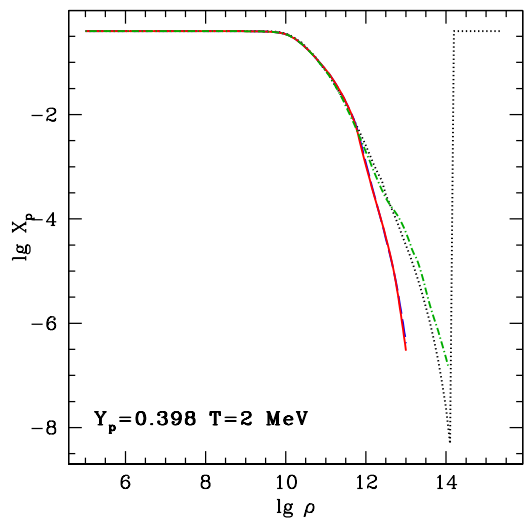}
\caption{\noindent  The mass fraction of free protons, $X_p$,
as a function of density
for $Y_{\rm e}=0.398$ and $T=2$ MeV.
The notation is the same as in Fig.~\ref{fig:1d-xap-y316t2}.
        }
\label{fig:1d-xap-y3986t2}
\end{figure}

We comment here on the difference from the calculated EOS
using the treatment  by  \citet{maz79}.
The calculation done without the second line in Eq. (\ref{eq:uznik}),
which express the dependence of binding energy of nucleus
on the nucleon fractions, are shown by dash-dotted lines
in Figs. \ref{fig:1d-xap-y316t2} to \ref{fig:1d-sn-y316t2}.
In general, the treatment without iterations gives similar
results to the one with iterations to calculate consistently
the modified binding energy due to the presence of nucleons
at this condition.
In Fig. \ref{fig:1d-xap-y3986t2}, we show the mass fraction
of free protons as a function of density
for $Y_{\rm e}=0.398$ and $T$=2 MeV.
We have found the treatment by \citet{maz79} provides
similar results to those of the Shen EOS and our treatment
gives lower proton fraction by large factors.
This may influence the electron capture rate in collapsing supernova cores.

We remark on the dependence of mass models.
The calculations by two mass models are similar to each other
and one can hardly see the difference.
Only in the prediction of mass number and proton number,
there are slight differences between the two lines.
In this respect, the derivation of bulk quantities,
such as entropy and pressure, can safely
be calculated without the uncertainties of mass formulae.
We note, however, that the detailed composition depends
on the adopted mass formula as we will show below.

\subsection{Nuclear composition}\label{sec:result-composition}

\begin{figure*}
\includegraphics[width=12cm]{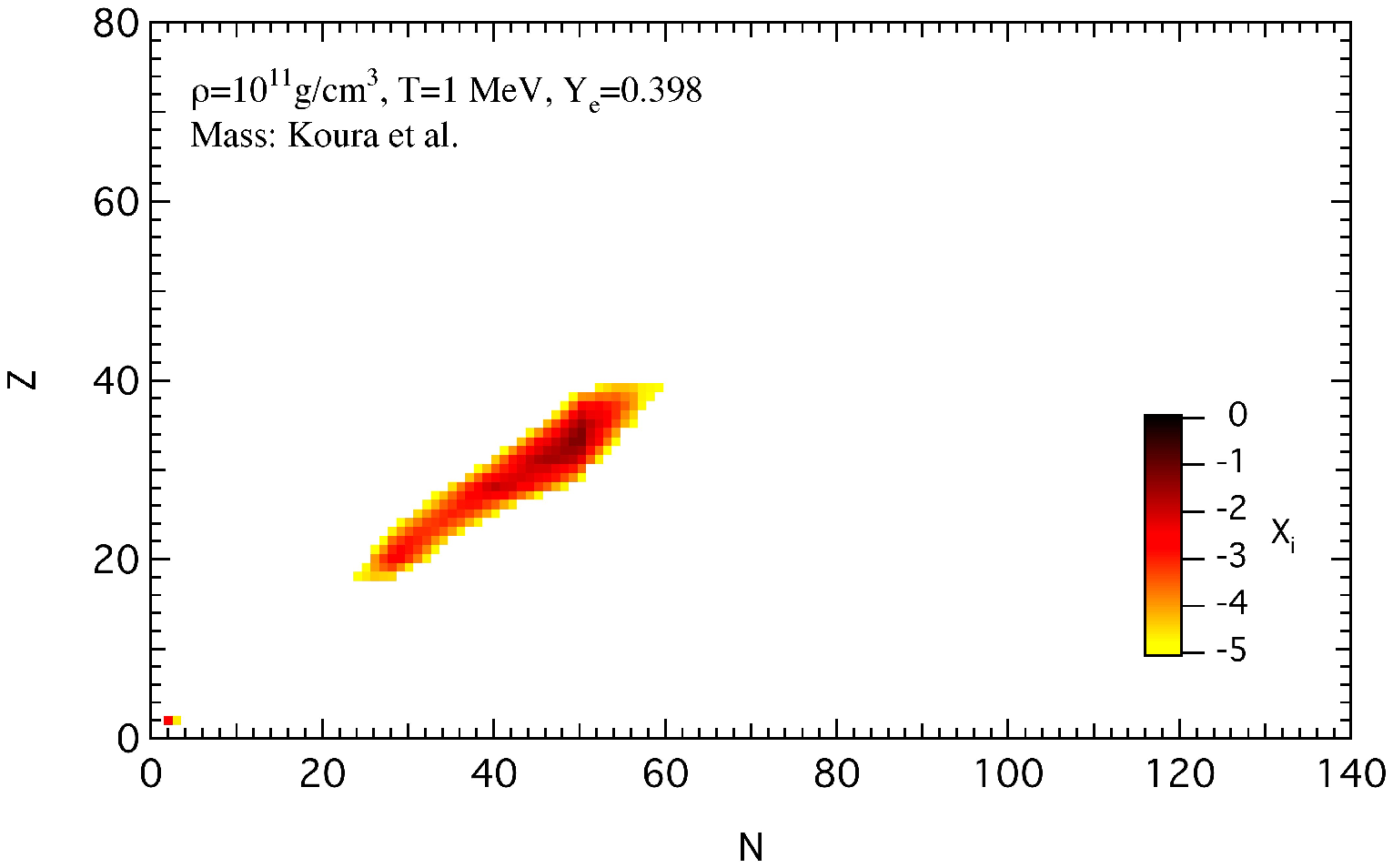}
\caption{\noindent  The mass fraction of nuclei, $X_A$, having
proton number, $Z$, and neutron number, $N=A-Z$
in the nuclear chart
for $\rho=10^{11} g/cm^3, Y_{\rm e}=0.398$ and $T=1$ MeV.}
\label{fig:koura-r11y398t1}
\end{figure*}

\begin{figure*}
\includegraphics[width=12cm]{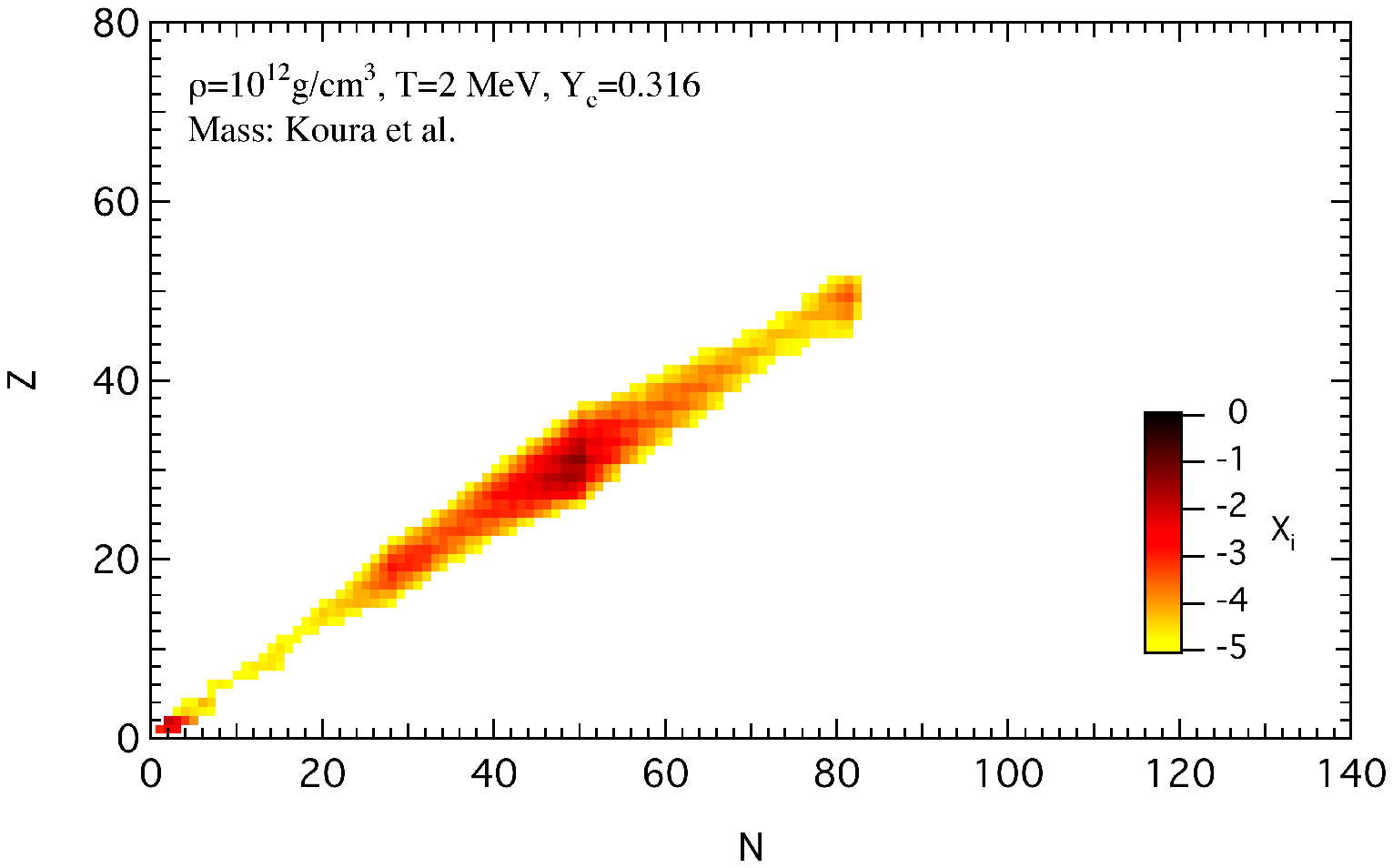}
\caption{\noindent  The mass fraction of nuclei, $X_A$, having
proton number, $Z$, and neutron number, $N=A-Z$
in the nuclear chart
for $\rho=10^{12} g/cm^3, Y_{\rm e}=0.316$ and $T=2$ MeV.}
\label{fig:koura-r12y316t2}
\end{figure*}

\begin{figure*}
\includegraphics[width=12cm]{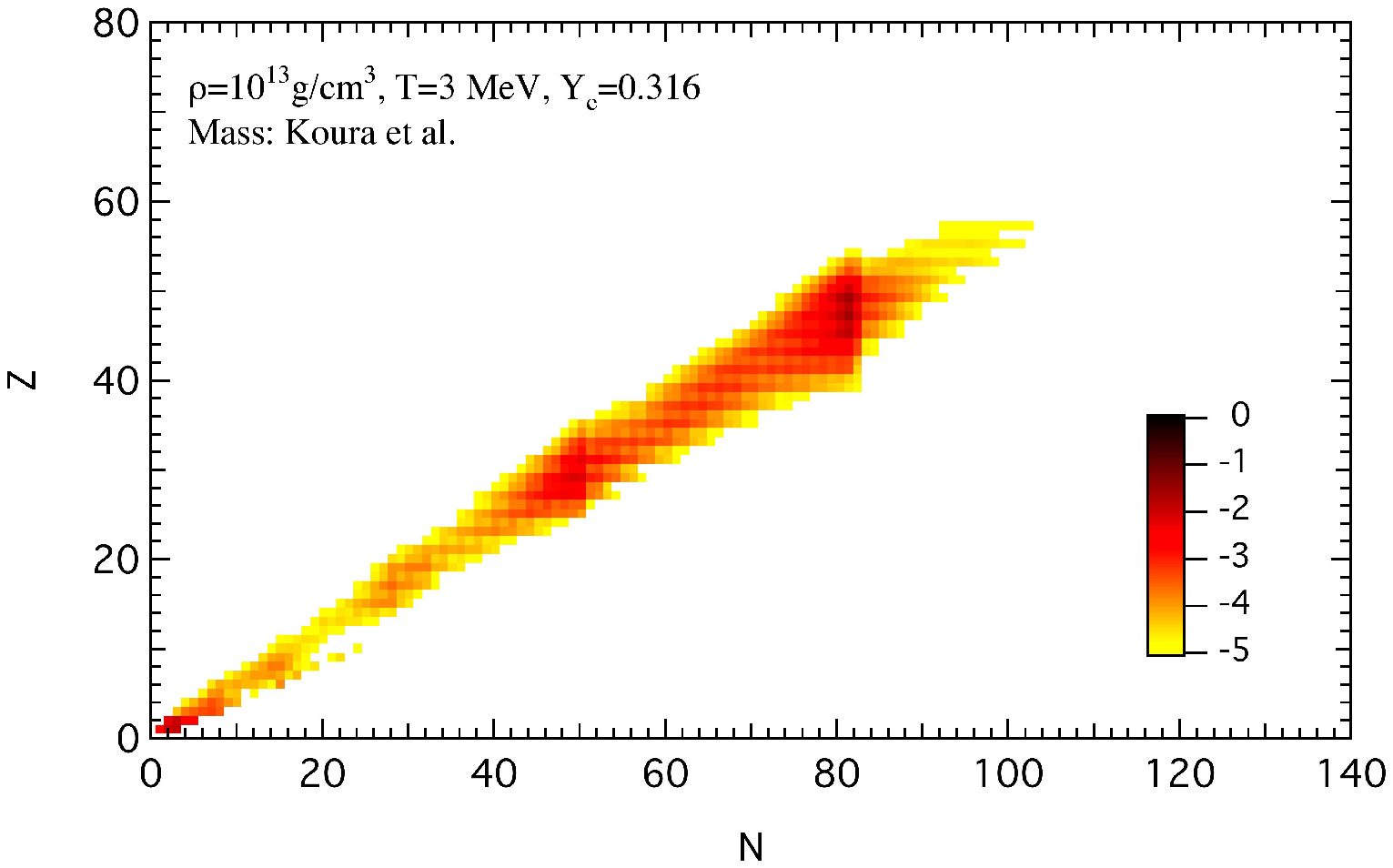}
\caption{\noindent  The mass fraction of nuclei, $X_A$, having
proton number, $Z$, and neutron number, $N=A-Z$
in the nuclear chart
for $\rho=10^{13} g/cm^3, Y_{\rm e}=0.316$ and $T=3$ MeV.}
\label{fig:koura-r13y316t3}
\end{figure*}

We explore the calculated composition using the mass formula by \citet{kou05}
in Figs. \ref{fig:koura-r11y398t1}, \ref{fig:koura-r12y316t2}
and \ref{fig:koura-r13y316t3} for three typical conditions
of supernova core.
We took the conditions from the numerical simulation of
core-collapse from a 15M$_\odot$ star by \citet{sum05}.

At $\rho$=10$^{11}$ g/cm$^{3}$, where electron captures goes on
during the gravitational collapse from the initial Fe core,
nuclei up to $A$ $\sim$ 100 appear mostly with the peak abundance
at $Z$ $\sim$ 34 and the magic number $N$=50.
By the time the density reaches $\rho$=10$^{12}$ g/cm$^{3}$,
electron capture is about to cease and $Y_{\rm e}$ becomes smaller.
Because of higher temperature due to the collapse,
the distribution becomes wider, reaching $A$ $\sim$ 132
at the double magic nuclei, $^{132}$Sn.
Neutron-richness (small $Y_{\rm e}$) makes the peak position
at more neutron-rich but at $N$=50.
At high density of 10$^{13}$ g/cm$^{3}$ and high temperature of 3 MeV,
which corresponds to the moment just before the core bounce,
the distribution extends from nucleons to $A$ $\sim$ 160 continuously.
We see the effect of magic numbers for neutron number 50 and 82
and even-odd numbers, which exist in nuclear mass data.

\begin{figure*}
\includegraphics[width=12cm]{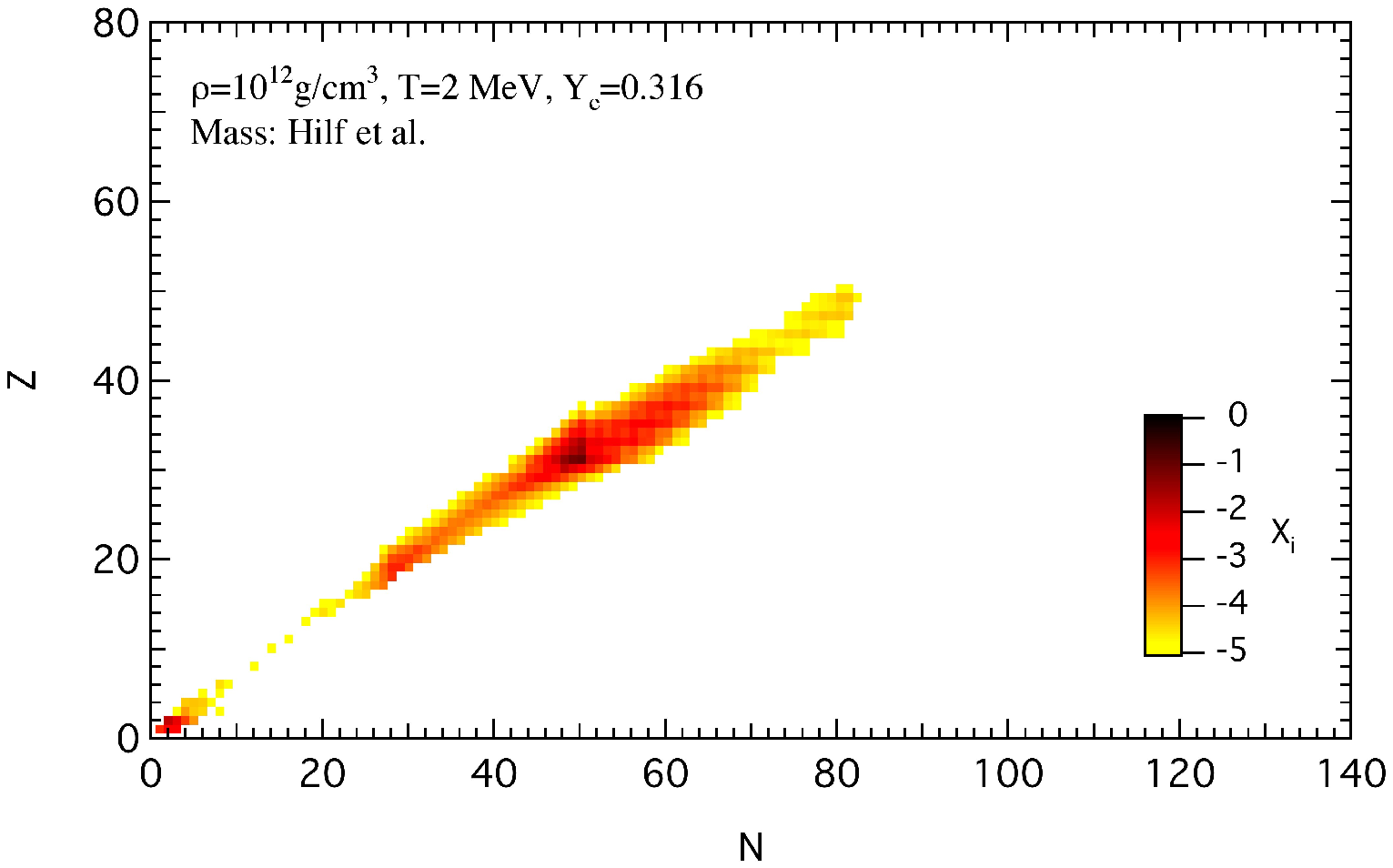}
\caption{\noindent  Same as Fig.~\protect\ref{fig:koura-r12y316t2} but for \citet{HilfGT76}}.
\label{fig:hilf-r12y316t2}
\end{figure*}

In order to demonstrate the difference of composition
due to the mass models, we show the case of $\rho$=10$^{12}$ g/cm$^{3}$
with the mass data by \citet{HilfGT76} in Fig. \ref{fig:hilf-r12y316t2}.
Through the comparison with Fig. \ref{fig:koura-r12y316t2},
the distribution is narrower both in $N$ and $A=N+Z$ directions.
The peak position is quite close each other in the two models.
We see similar differences also in the case of $\rho$=10$^{11}$ g/cm$^{3}$
and $\rho$=10$^{13}$ g/cm$^{3}$, especially on the strength of magic numbers.
Neutron-richness at high density causes larger differences among them.

\subsection{Difference in Coulomb corrections}\label{sec:result-Coulomb}

\begin{figure}
\includegraphics[width=0.45\linewidth]{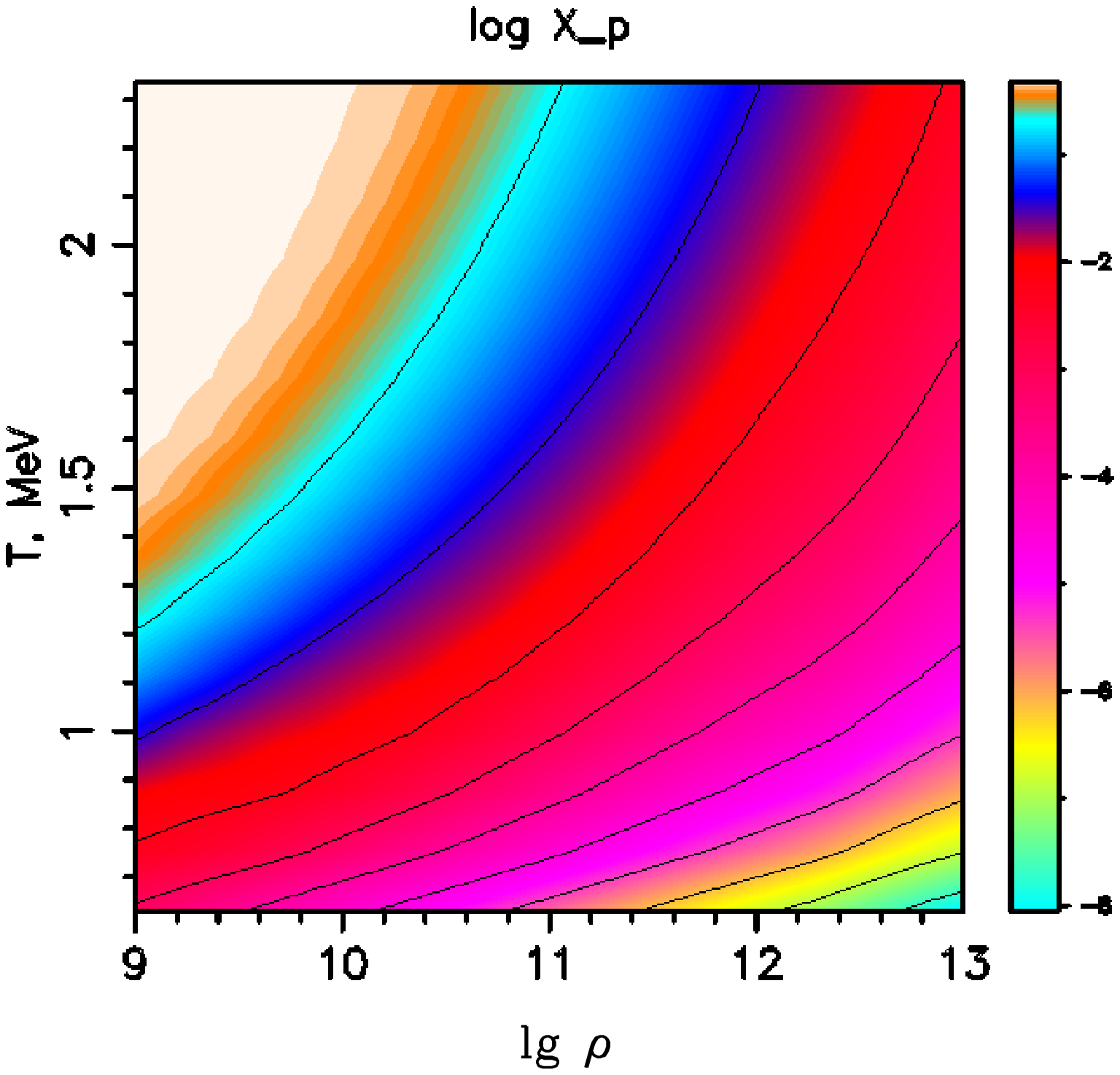}
\includegraphics[width=0.45\linewidth]{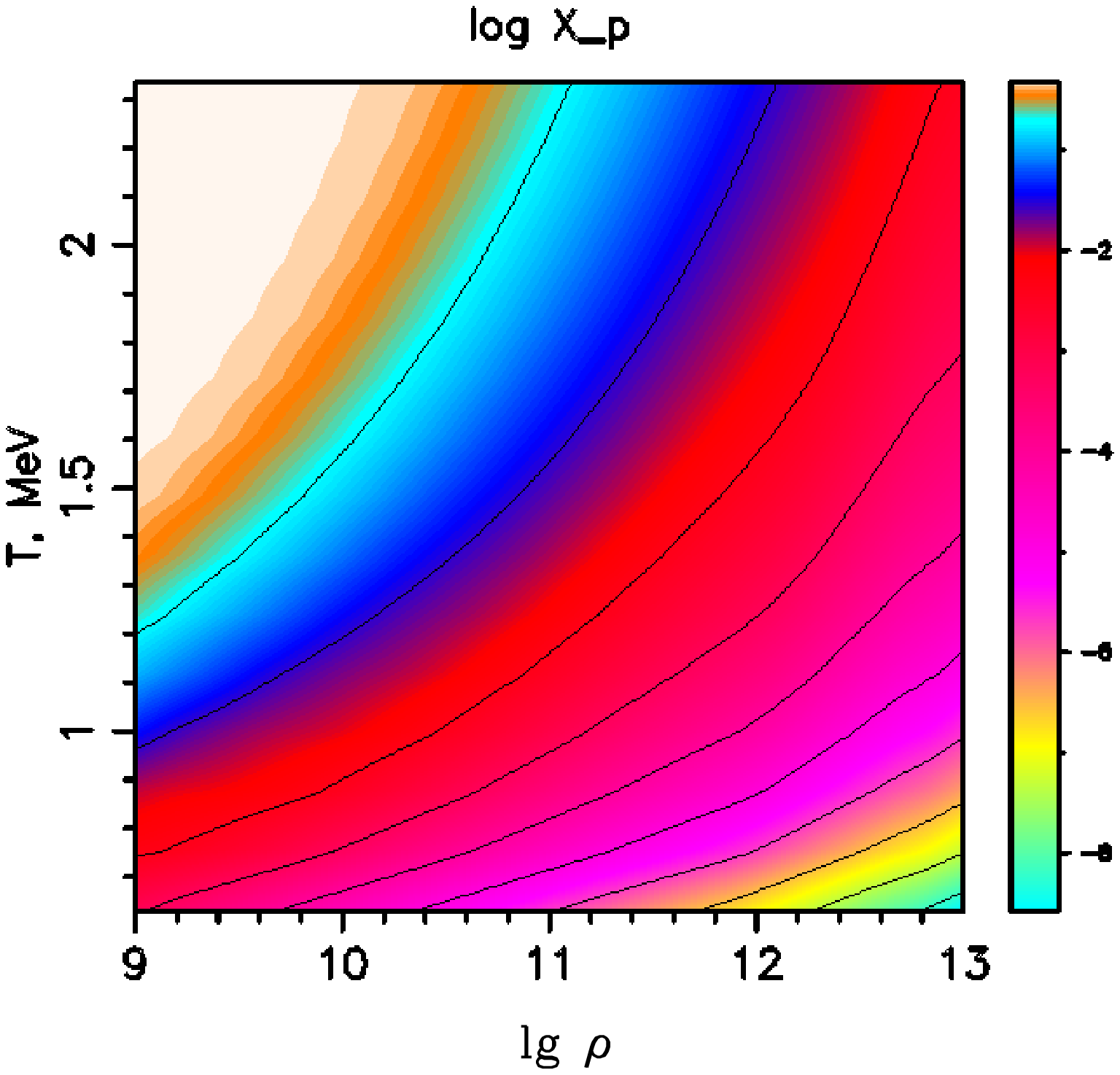}
\caption{\noindent  The isocontours of
the  proton fraction, $X_p$,
for $Y_{\rm e}=0.473$. Left panel: the Coulomb correction is taken
as in Eq.(\ref{eq:zanoza}). Right panel: as in Eq.(\ref{eq:gusj})
        }
\label{fig:2d-xp-y473}
\end{figure}

\begin{figure}
\includegraphics[width=0.45\linewidth]{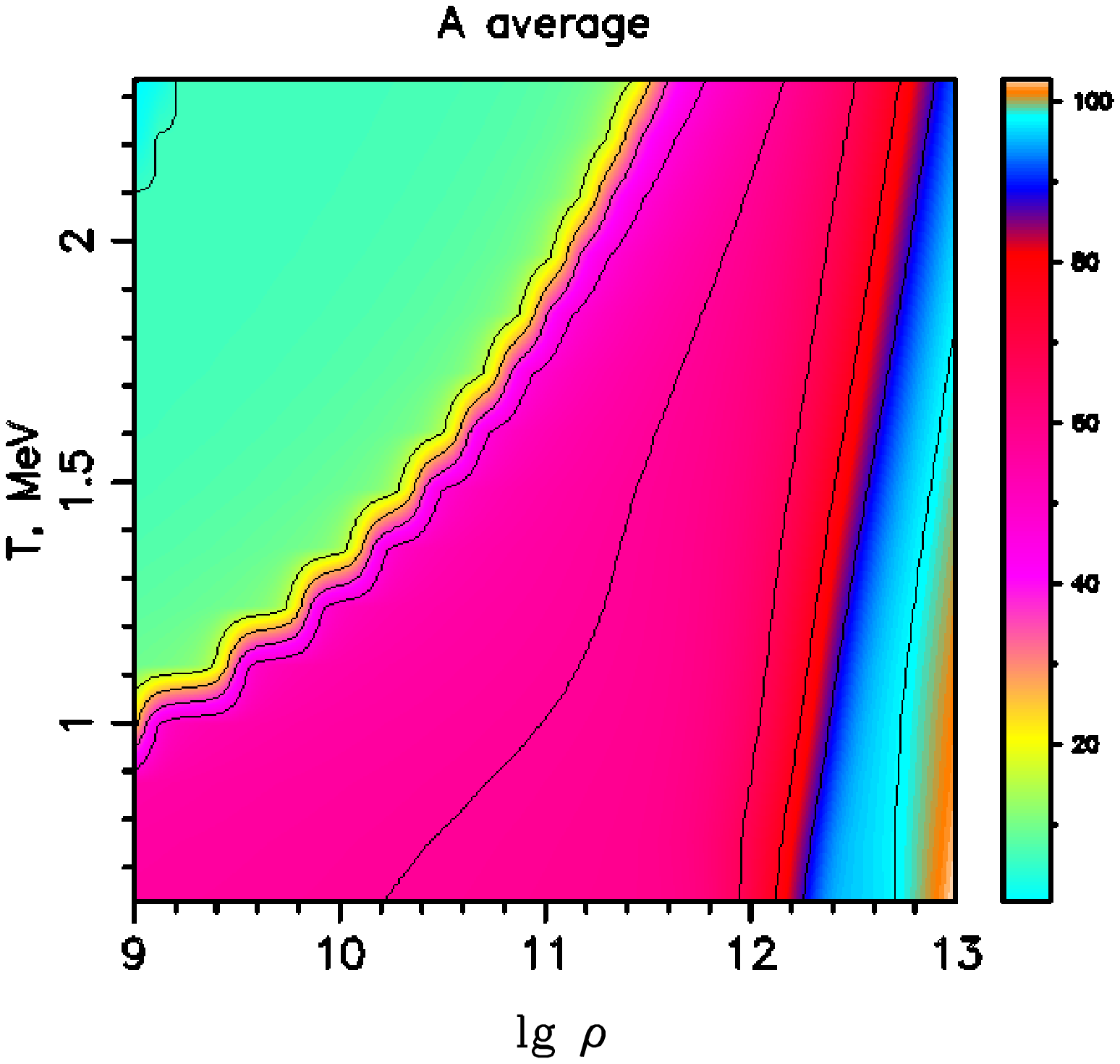}
\includegraphics[width=0.45\linewidth]{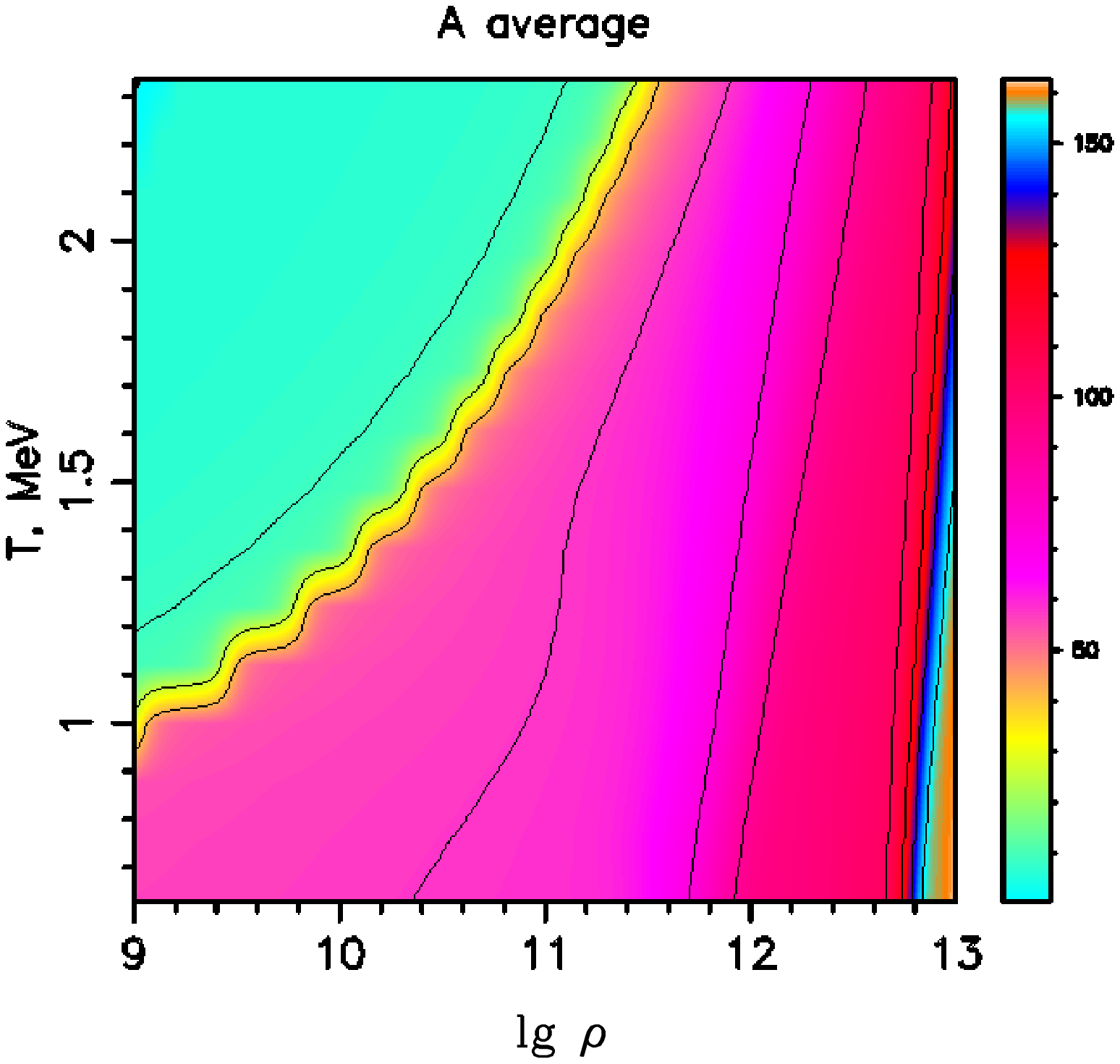}
\caption{\noindent  The isocontours of
the average mass number, $A$, of heavy nuclei,
for $Y_{\rm e}=0.473$. Left panel: the Coulomb correction is taken
as in Eq.(\ref{eq:zanoza}). Right panel: as in Eq.(\ref{eq:gusj})
        }
\label{fig:2d-a-y473}
\end{figure}

As we have discussed above, there is some controversy on the form of Coulomb
corrections in the literature.  We compare in  Fig.\ref{fig:2d-xp-y473}
the effect of changing corrections as in formula (\ref{eq:zanoza}) and in
(\ref{eq:gusj}) on the proton fraction. The effect is visible only at highest density
but still very small. A bit more pronounced effect is obvious in Fig.\ref{fig:2d-a-y473}
for the average mass of heavy nuclei again at highest density when the NSE model
is near the border of applicability. However, a more detailed study is
needed here along the lines undertaken recently by \citet{nad05}.

\section{Summary and discussions}\label{sec:summary}

We have studied the properties of hot and dense matter
at the conditions relevant to supernova cores
in the extended Saha approach.
The multi-composition of nuclei is taken into account
by solving the chemical equilibrium among nuclei, neutrons
and protons.
The contributions to chemical potentials
due to the degeneracy of nucleons, Coulomb effects,
nuclear effects, surface effects are included
in the Saha formulation.
The modification of binding energies of nuclei
due to the existence of nucleons outside nucleus
is solved self-consistently.
This is different from the treatment of \citet{maz79}.
Cf. \cite{nad05} were the difficulty of self-consistent treatment is pointed out.
We have adopted the recent mass formula by \citet{kou05},
covering $\sim$20000 nuclides,
and compared the results with the ones by \citet{HilfGT76}.

We have calculated extensive grids 
of density
and temperature for several choices of $Y_{\rm e}$.
We found that the obtained quantities behave in general
very smoothly.
When we compared with the values in the Shen EOS,
which is a popular set of EOS for supernova simulations,
they coincide with each other closely for thermodynamical
quantities.
This may suggest that the one species treatment of nuclei
adopted in most of modern supernova EOS's is a good
approximation in this regard.
However, the appearance of nuclei starts at lower
density than in the Shen EOS and the mass fractions
of alpha particles and nuclei are different in two codes.
The average mass and proton numbers of nuclei are larger
at low density and smaller at high density
than those obtained from the Shen EOS.
The former arises from the appearance of light
nuclei around $A\sim 10$ in addition to alpha particles which are only included
in Shen EOS.
We explore also the composition in supernova core
in the nuclear chart.
We found that the distribution extends widely for
high density and temperature
and it depends on the choice of mass formula,
which provides different shell effects.

Our approach  is rather simple and may be compared
with more sophisticated modern treatment of nuclear reactions
based on the statistical multi-fragmentation model
(SMM): see \cite{Botvina08}.
Recently an exploration on the difference between the
treatments under single and multi nuclear species approximations
with phenomenological mass formulae has been carried out to
demonstrate the overestimation of mass number in the single
nuclear species \cite{Souza08}.

The current finding assists understanding of
the gravitational collapse of massive stars
for supernova explosions.
The difference of composition from the one species
treatment in the Shen EOS suggests that the proper
treatment in multi-composition of nuclei is important
and may significantly affect the dynamics of core
collapse.
Since the electron capture on nuclei and free protons
is crucial to determine the size of bounce core,
i.e. the location of initial launch of shock wave, one should take
into account the composition properly.
Such efforts have been made to evaluate the total
electron capture rates on nuclei by averaging
the distributions \citep{hix03}.
One should, however, work carefully to predict
the composition, since the mass  and the nuclear
effects may affect the detailed abundance.
It is also necessary to calculate the equation of state,
thermodynamical quantities and compositions in a consistent manner.
We are aiming to apply the extended Saha treatment
to provide EOS sets for supernova simulations
in future.

\section*{Acknowledgements}

A part of this work was done while S.~Blinnikov, I.~Panov, and K.~Sumiyoshi
were visiting Max Planck Institut f{\"u}r Astrophysik, Garching.
They are especially thankful to W.~Hillebrandt, E.M\"uller and H.-Th. Janka
for the hospitality and to WH and EM for reading the manuscript at the early and final stages of the work. SB cordially thanks K.Nomoto, K.Sato, and H.Murayama for hospitality at RESCEU and IPMU in Tokyo where he could complete the paper.
His work is supported by World Premier International Research
Center Initiative (WPI), MEXT, Japan.
We are grateful to D.Yakovlev for drawing our attention to
the problem of Coulomb corrections
in nuclear interactions and for his valuable consultations.
Late C.Engelbrecht has given us his results on partition functions prior to publication.
The work in Russia is supported partly by the grant RFBR 07-02-00830-a
and by Scientific School Foundation under grants 2977.2008.2,
3884.2008.2, and by a grant IB7320-110996 of the Swiss National Science
Foundation.
KS is partly supported by the Grants-in-Aid for the
Scientific Research (18540291, 18540295, 19104006, 19540252, 20105004) in Japan.
SB is grateful to Elena Blinnikova for typing formulae in \TeX,
and to Mona Frommert for her valuable help at one of the critical stages of the project.

\bigskip


\newpage



\end{document}